%% file: ms.tex
\documentclass[9pt,sigconf]{acmart}

\renewcommand\footnotetextcopyrightpermission[1]{} 
\usepackage{subfiles}
\usepackage{xcolor}
\usepackage[color=white]{todonotes}
\usepackage{amsmath}
\usepackage[caption=false]{subfig}
\DeclareMathOperator*{\argminB}{argmax}

\usepackage{booktabs}
\usepackage{multirow}
\usepackage{soul}

\usepackage{caption}

\newcommand{\Sys}{Wi-Fringe}

\definecolor{dkgray}{rgb}{0.4,0.4,0.4}
\definecolor{dkgreen}{rgb}{0,0.6,0}
\definecolor{gray}{rgb}{0.5,0.5,0.5}
\definecolor{mauve}{rgb}{0.58,0,0.82}

\newcommand{\mytt}[1]{{\small \texttt{#1}}}
\newcommand{\rnote}[1]{\textcolor{black}{#1}}
\newcommand\gnote[1]{\textcolor{black}{#1}}

\newcommand\bnote[1]{\textcolor{black}{#1}}

\author{}

\begin{document}

\title{\Sys: Leveraging Text Semantics in WiFi CSI-Based Device-Free Named Gesture Recognition}

\author{Md Tamzeed Islam}
\affiliation{%
  \institution{UNC Chapel Hill}
}
\email{tamzeed@cs.unc.edu}

\author{Shahriar Nirjon}
\affiliation{
\institution{UNC Chapel Hill}
}
\email{nirjon@cs.unc.edu}

\input{tex/01_Abstract.tex}

\maketitle

\input{tex/02_Introduction_local.tex}

\input{tex/03_usecase_local.tex}

\input{tex/05_background_new.tex}

\input{tex/04_design_short.tex}

\input{tex/06A_representation.tex}

\input{tex/06B_CMP.tex}
\input{tex/06C_classify.tex}

\input{tex/07_implementation.tex}

\input{tex/08_evaluation.tex}
\input{tex/10_discussion.tex}

\input{tex/09_literature.tex}

\input{tex/11_conclusion.tex}

\bibliographystyle{ACM-Reference-Format}

\end{document}

%% file: tex/01_Abstract.tex
\begin{abstract}
The lack of adequate training data is one of the major hurdles in WiFi-based activity recognition systems. In this paper, we propose \Sys, which is a WiFi CSI-based device-free human gesture recognition system that recognizes \emph{named} gestures, i.e., activities and gestures that have a semantically meaningful name in English language, as opposed to arbitrary free-form gestures. Given a list of activities (only their names in English text), along with zero or more training examples (WiFi CSI values) per activity, \Sys is able to detect all activities at runtime. In other words, a subset of activities that \Sys detects do not require any training examples at all. This is achieved by leveraging prior knowledge of these activities from another domain, i.e., text. We show for the first time that by utilizing the state-of-the-art semantic representation of English words, which is learned from a massive dataset such as the Wikipedia (e.g., Google's word-to-vector~\cite{mikolov2013efficient}) and verb attributes learned from how a word is defined (e.g, American Heritage Dictionary), we can enhance the capability of WiFi-based named gesture recognition systems that lack adequate training examples per class. We propose a novel cross-domain knowledge transfer algorithm between radio frequency (RF) and text to lessen the burden on developers and end-users from the tedious task of data collection for all possible activities. To evaluate \Sys, we collect data from four volunteers for 20 activities in two environments. 
We show that \Sys achieves an accuracy of up to $90\%$ for two unseen activities and $61\%$ for up to six unseen activities.
\end{abstract}


%% file: tex/02_Introduction_local.tex
\section{Introduction}

\label{sec:intro}

The widespread adoption of WiFi in indoor spaces, and the accessibility of WiFi signal characteristics such as the signal strength (RSSI) and channel state information (CSI) in commodity WiFi chipsets, make WiFi an attractive technology for human activity monitoring. Alternate solutions that use wearables like smartwatches and activity trackers are becoming less attractive due to their usage adherence issue, and systems that use cameras to monitor home activities raise serious privacy concerns. WiFi sensing, on the other hand, is device-free, non-intrusive, and less privacy-invasive. Hence, in recent years, we have seen an increase in WiFi-based sensing and inference systems whose feasibility has been demonstrated in application scenarios such as home activity monitoring~\cite{wang2014eyes}, sleep monitoring~\cite{liu2014wi}, gesture-based smart device interaction~\cite{abdelnasser2015wigest, adib2013see}, and health vitals tracking~\cite{liu2016contactless}. 

 \begin{figure}[!thb]
	\centering
	\includegraphics[width = 0.45\textwidth]{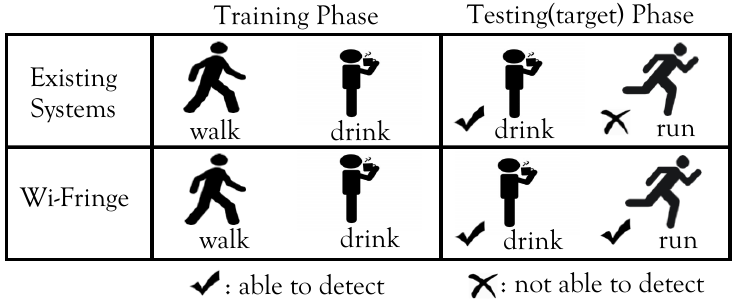}
	\caption{Unlike existing systems, \Sys is able to recognize \textit{run} in the testing phase, even though it did not see any training example of \textit{run} in the training phase. }
	\label{fig:intro}
\end{figure}

A vast majority of WiFi-based activity recognition systems employ either template matching algorithms or machine learning classifiers --- ranging from traditional classifiers like support vector machines~\cite{wang2015robust} to advanced deep convolutional neural networks~\cite{yue2018extracting}. These algorithms require a decent number of training examples for each class of activity in order for the system to accurately model them. Furthermore, the capability of these systems are fundamentally limited by the number of activity classes for which the system has been trained for. When these systems are presented with a completely new type of activity, there is no built-in mechanism to make an educated guess about the possible class label for that unseen example. 

Figure~\ref{fig:intro} illustrates this scenario. When a system is trained to recognize only \mytt{\{walk, drink\}}, but is presented with an example of an unseen activity, e.g., \mytt{run}, it is likely to detect the activity as either \mytt{walk} (based on the closest match) or it will determine that it is an unknown category (based on a distance threshold). At present, existing systems have no inherent mechanism to infer that the activity could be \mytt{run}, as these systems have no prior knowledge of how an activity called \mytt{run} might be.         

A na\"ive way to deal with the above problem is to train a system with examples of all possible activities that it may ever encounter. However, this is not feasible for several practical reasons. First, despite recent efforts in environment-invariant activity classification~\cite{jiang2018towards}, existing WiFi-based sensing systems are prone to environment changes \bnote{due to RF signals' dependency on surroundings for reflection, refraction and scattering.} Hence, it puts the burden on the end-users to provide the training data --- which is time-consuming, error-prone, and in general, an inconvenience. Second, even with environment-invariant techniques, due to the diversity of human activities, it is not feasible to build models for all possible activities for users of all demographics. This motivates us to devise a WiFi-based sensing system that recognizes activities and gestures without prior examples.

In this paper, we propose the first system, called the \emph{\Sys}, which can infer activities from WiFi data without requiring prior training examples for all of its activity classes. The principle behind \Sys is popularly known as the \emph{zero-shot learning}~\cite{lampert2014attribute,frome2013devise}, which is an active research topic in computer vision and image classification fields. \bnote{Recently, zero-shot learning has also been applied to general-purpose sound recognition problem~\cite{islam2019soundsemantics}. These techniques are not directly applicable to RF-based gesture recognition problems, since gestures require tracking sequential properties of the signal and external knowledge about the attributes that defines an activity.} To the best of our knowledge, we are the first to apply zero-shot learning in RF-based device-free activity recognition problem. The core idea of zero-shot learning is to exploit information or learned knowledge from other sources such as textual descriptions, rules, and logic. For example, to teach the concept of \mytt{run} to a system that has already learned to recognize \mytt{walk}, instead of training it with many examples of \mytt{run}, we can add a rule into the system, e.g., ``\mytt{run} is just like \mytt{walk} but it's 3 to 5 times faster.'' At runtime, the system will use this additional information to classify a \mytt{run} activity correctly. 

In \Sys, to embed such rules between \emph{seen} classes (i.e., explicitly trained) and \emph{unseen} classes (i.e., not explicitly trained) in an RF sensing system, we exploit attributes and context-aware representation of English words as the additional source of knowledge. Through an advanced RF-domain to text-domain projection algorithm, we blur the difference between an activity's RF signature and its corresponding English word/phrase by representing them in the same vector spaces, i.e., \emph{word embedding}~\cite{mikolov2013distributed} and \emph{word attribute}~\cite{zellers2017zero} spaces. We exploit the attributes and semantic relationship between English words as the background knowledge to classify an activity from WiFi that the system has never seen before. We combine these algorithmic steps to develop an advanced activity recognition system that we name---\Sys, as it can detect previously unseen \emph{fringe} activities beyond what it is trained for. 


The intuition behind \Sys is that the WiFi signature of an activity correlates with the corresponding verb's semantic and attribute information. Like two similar activities perturb the WiFi signals similarly, when we describe these two activities in English sentences, we see a similar likeness between the sentences. We generalize this notion for an arbitrary number of activities represented in both RF and text domains, and strive to find a projection between the two representations from the RF domain to the text domain. By learning this projection, we gain the ability to find the corresponding English word from the RF representation of any arbitrary activity. 



Projecting RF signals onto the space of textual representation is non-trivial and poses several challenges that are addressed in this paper. First, we propose the very first context-aware RF features by explicitly learning the transition of \textit{states} (i.e., micro-activities) in an activity . We show that such a representation is robust and yields better features for activity recognition in general. Second, we propose a neural network architecture to merge text- and RF-domain representations of activities so that WiFi CSI data are mapped to the attributes and distributional characteristics of English words. This results in the first \emph{cross-modal RF embedding} work, and paves the way for device-free WiFi-based activity classification without requiring training data for all activities. Third, we propose a two-level classifier that is capable of classifying both \emph{seen} and \emph{unseen} activity types. This makes \Sys a generalized system for classifying a wide variety of human activities. 

We develop \Sys using Intel Network Interface Card (NIC) 5300~\cite{nic-card} which captures the WiFi CSI data. To develop the machine learning models, we collect training and testing data from four volunteers in a multi-person apartment as well as in an office building for \bnote{20} activity classes --- which is, to the best of our knowledge, the largest collection of activities used in any WiFi-based device-free gesture recognition system till date. \bnote{We show that \Sys is able to recognize activities with 62\%-90\% accuracy, as we vary the number of unseen classes from six to two.}


The contributions of the paper are the following:
\begin{itemize}
     \item We describe \Sys, which is the first system that addresses the problem of learning without examples for WiFi based device-free \gnote{named} activity recognition.
     
     \item We propose a new approach to learn state-aware RF representation  to preserve the transitional characteristics of states in an activity. This representation by itself improves the classification performance in supervised learning.    
     
     \item We propose the first algorithm that exploits semantic knowledge and attributes in text to enable classification of activities from WiFi data without prior training examples.
     
     \item We collect data from four volunteers, from two environments, for 20 activities, and show that \Sys beats state-of-the-art WiFi-based activity recognition algorithms by $30\%$.
     
 \end{itemize}

%% file: tex/03_usecase_local.tex
\section{Example Usage}

\label{sec:apps}

\Sys relieves the developers and end-users from the burden of extensive data collection and training and is able to classify new types of activities for which no training examples are provided. 

There are many WiFi-based activity recognition systems such as elderly monitoring~\cite{palipana2018falldefi}, home activity recognition~\cite{wang2014eyes}, gesture based media controller~\cite{abdelnasser2015wigest}, and gesture based gaming~\cite{he2015wig} where the use \Sys can significantly improve the performance as well as the intelligence of an application. We describe one such scenario to illustrate the capability and usage of \Sys.

\begin{figure}[!htb]
	\centering
	\includegraphics[width = 0.45\textwidth]{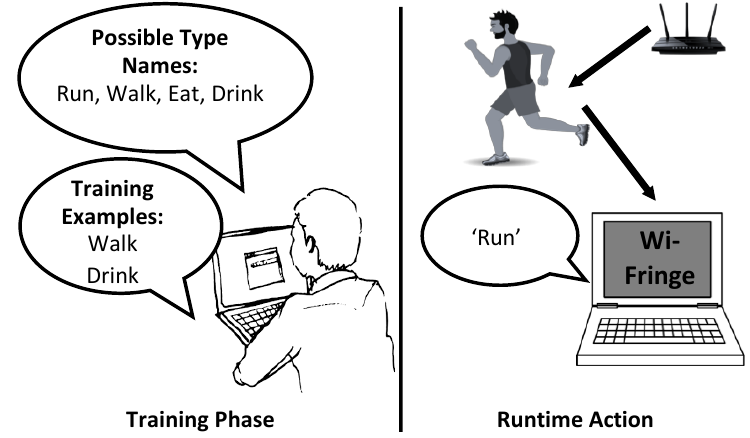}
	\caption{\Sys is able to recognize activities for which is has not been explicitly trained for.}
	\label{fig:usecase}

\end{figure}





Consider a simplified home activity recognition scenario which is shown in Figure~\ref{fig:usecase}. The user of \Sys wants to monitor four activities: \mytt{\{run, walk, eat, drink\}}. At first, they input this list of four activities (only the names) to \Sys. \Sys then asks the user to provide zero or more training examples for each activity. However, our user provides training examples for two out of the four classes: \mytt{\{walk, drink\}} and does not provide any example for the remaining two classes: \mytt{\{run, eat\}}. 

Despite the lack of training examples, \Sys is able to recognize the \mytt{run} activity at runtime by combining its knowledge of \mytt{walk} (from training) with its knowledge of the relationship between \mytt{run} and \mytt{walk} (which it acquired from external sources such as textual embedding).

%% file: tex/05_background_new.tex
\section{Background}
\label{sec:back}

\subsection{Channel State Information (CSI)} 


In wireless communication, a transmitter and a receiver talk to each other using a certain frequency or a range of frequencies, which is often called a \textit{channel}. The \textit{Channel State Information} (CSI), as the name implies, describes the properties of a wireless channel. For example, when wireless signals travel through the medium, they fade, they get reflected and scattered by obstacles on the way, and their power decays with the distance traveled. The CSI is a measure of all these phenomena of a wireless channel. Mathematically, using the frequency domain terms, we express the relationship between the transmitted signal $X(f, t)$, the channel frequency response (CFR) $H(f, t)$, and the received signal $Y(f, t)$ as: $Y(f, t) = H(f, t)\cdot X(f, t) + N(f,t)$, where $N(f,t)$ denotes the noise. The CSI comprises of the CFR values, i.e., $\{H(f, t)\}$.

Data communication over a WiFi channel happens by sending the data bits simultaneously over 64 distinct frequencies (called the \emph{sub-carriers}) in parallel. From the WiFi Network Interface Card (NIC)~\cite{nic-card}, it is possible to obtain the CSI values of these subcarriers. The frequency response, $H(f,t)$ of each sub-carrier is a complex number, where the real and complex parts represent the amplitude and phase response, respectively. We only use with the real part (i.e., amplitude) in this paper as we find the phase values too noisy to be useful. For $N_{TX}$ transmitting antennas, $N_{RX}$ receiving antennas, and $N_S$ sub-carriers, we get a CSI matrix of complex numbers having the dimensions of $N_{TX}\times N_{RX}\times N_s$.

\subsection{Word Embedding}

The process of \emph{Word Embedding}~\cite{mikolov2013distributed} maps words in a natural language to vectors of real numbers in a manner that words that are commonly used in the same textual context are positioned closely in the vector space. Intuitively, word embedding expresses the distributional semantics of words with the motto that \emph{``a word is characterized by the company it keeps''}~\cite{firth1957synopsis}. Word embedding extracts distributional similarities among the words from a large scale text data, such as the Wikipedia, by observing the words that appear with similar words which define their context. For example, consider the words: \mytt{love} and \mytt{adore}. Syntactically these two words are quite different, but they often appear in similar semantic contexts, i.e., with similar words. Hence, the word embedding process would map these two words to two vectors whose distance is relatively closer than the embedding of two random words.   


\begin{figure}[!htb]
	\centering
	\includegraphics[width = 0.45\textwidth]{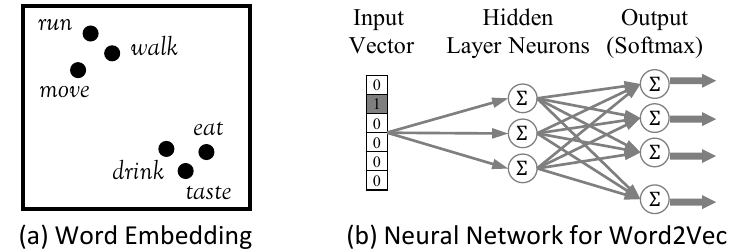}
	\caption{(a) Semantically similar words appear close to each other in the word embedding space. (b) A two layer neural network is used for Word2Vec~\cite{mikolov2013efficient} word embedding.}
	\label{fig:w2vec}
	\vspace{-0.5em}
\end{figure}

In Figure~\ref{fig:w2vec}(a), we plot the embedding of six English words. We only show the first two principal components to be able to visualize the vector representation in 2D. We observe that words with similar semantic meaning are closer in the word embedding space, e.g., \mytt{\{run, walk, move\}}, whereas words having different semantic meaning are far, e.g., \mytt{run} and \mytt{eat}. In Natural Language Processing (NLP), researchers use techniques such as neural network~\cite{wu1992introduction} and co-occurrence matrix~\cite{partio2002rock} to project words onto a $k$-dimensional vector space, and words appearing in similar textual contexts get closer embedding in the vector space. One of the most popular method to extract word embedding from a large scale corpus is \textit{Word2Vec}~\cite{mikolov2013efficient}, which uses a two layer neural network to predict the surrounding words of a target word. 

Figure~\ref{fig:w2vec}(b) shows a two-layer neural network, where a word, represented by its one-hot encoding~\cite{turian2010word}, is the input and the last layer of the network outputs the probability of other words' being its neighbour. Word-pairs of the form $\{(word, neighbor)\}$ are used to train this network, which explicitly learns to predict neighbour words of a given word, and in the process, the network implicitly learns the context of the words. After the training phase, the output of the hidden layer is used as the embedding that projects an input word to a low-dimension vector space.

\subsection{Attribute Embedding}




While word embedding captures the co-occurrence information of words used in the same context, it does not quite describe the meaning of a word. Recently, natural language processing community has proposed an effective method to learn the attributes of English \emph{verbs} from their dictionary definitions~\cite{zellers2017zero}. In this new method, verbs are expressed in terms of a set of attributes. Each verb is expressed as a vector of real numbers where each element of the vector corresponds to an attribute.    



\begin{table}[!htb]
    \centering
    \resizebox{.475\textwidth}{!}{
    \begin{tabular}{r|l}
        \textbf{Word} & \textbf{Representation} \\ \hline   
        Drink &  Dictionary:  ``To take into the mouth and swallow a liquid.''\\
        & Attributes:  \textit{(Motion, Social, Object, Head, \dots) = (low, solitary, true, true, \dots)}\\

        Sip &  Dictionary:  ``To drink in small quantities.''\\
        & Attributes:  \textit{(Motion, Social, Object, Head, \dots) = (low, social, true, true, \dots)}\\
        
        Drool &  Dictionary:  ``To let run from the mouth.''\\
        & Attributes:  \textit{(Motion, Social, Object, Head, \dots) = (none, solitary, false, true, \dots)}\\        
    \end{tabular}
    }
    \caption{Word definitions and attributes.}
    \label{tab:word-attr}
	\vspace{-1.55em}

\end{table}

Table~\ref{tab:word-attr} provides a simplified example. Three verbs: \emph{Drink}, \emph{Sip}, and \emph{Drool} are expressed in terms of four attributes: \emph{Motion}, \emph{Social}, \emph{Object}, \emph{Head}, where the attributes correspond to the degree of motion, degree of social engagement, use of objects, and use of head, respectively. We intentionally left the list of attributes open to emphasize that additional attributes are necessary to encode the dictionary definitions of a large number of English verbs.  




The process of attribute extraction is a standard supervised learning task where the attributes are predicted from a word's dictionary definition. In~\cite{zellers2017zero}, 24 attributes (inspired by Linguistics and verb semantics) were chosen to represent a total of 1700 English verbs. The overall process has four steps. First, the \emph{Wordnik} API~\cite{word_link} is used to obtain the dictionary definitions of the verbs. Second, a deep neural network called the \emph{Bidirectional Gated Recurrent Unit} (BGRU)~\cite{zellers2017zero} is used to encode the dictionary definitions. Third, the word embedding of the verb is concatenated to the output of the BGRU to incorporate contextual information. Finally, the concatenated encoding is fed to another neural network to obtain the attribute embedding of the verb.      




%% file: tex/04_design_short.tex
\section{\Sys System Design}
 \Sys takes a short-duration WiFi CSI stream (e.g., 5--8 seconds) and a list of possible activity types (i.e., a list of tags) as the input, and processes the CSI stream through a signal processing pipeline to classify it as one of those given activity types. The design of \Sys is modular. Computationally expensive modules such as the one-time offline training of the classifiers are run on a server, while the end-to-end activity classification pipeline---from sensing to classification---is runnable on embedded systems such as smartphones and tablets\footnote{Recent developments~\cite{Schulz:WiNTECH:2017} have shown how to extract CSI on smartphones. In this paper, we conduct experiments using an Intel NUC~\cite{nuc-pc}.}.

\begin{figure}[!htb]
    \centering
    \includegraphics[width=0.49\textwidth]{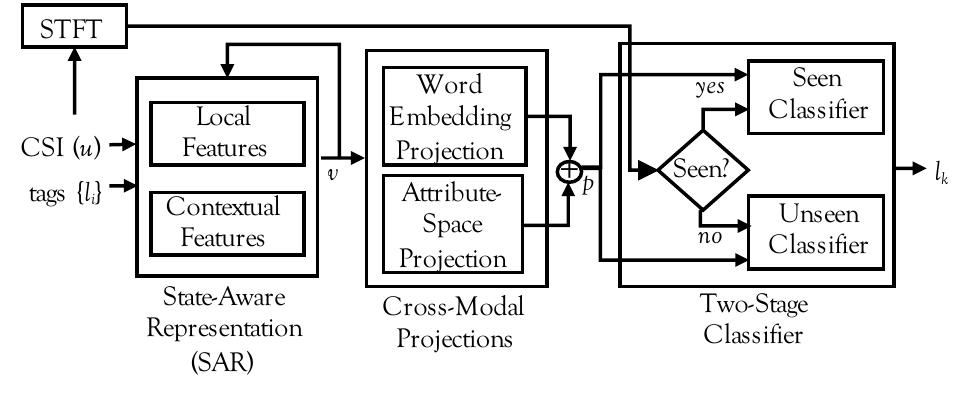}
       
    \caption{\Sys signal processing pipeline. }
    \label{fig:overview}
    \vspace{-0.5em}

\end{figure}

Figure~\ref{fig:overview} shows the signal processing pipeline of \Sys, which consists of three main steps: \emph{State-Aware Representation}, \emph{Cross-Modal Projections}, and \emph{Two-Stage Classifier}. The State-Aware Representation step extracts local and contextual features from the CSI stream. These features are projected onto the word and attribute spaces to incorporate external knowledge from the text domain in the Cross-Modal Projections step. The two-stage classifier determines if the input belongs to a seen or an unseen class and then classifies it accordingly.

There are two classes of activities that \Sys may encounter at runtime: \emph{seen} and \emph{unseen} classes. The \emph{seen} class refers to those activity types for which \Sys has labeled CSI streams for training. The \emph{unseen} class, on the other hand, refers to activity types for which \Sys does not have any training CSI stream. For example, in the home activity monitoring scenario of Figure~\ref{fig:usecase}, \mytt{\{walk, drink\}} are seen classes and \mytt{\{run, eat\}} are unseen classes.

The next three sections describe the algorithmic details of these three components of \Sys.

%% file: tex/06A_representation.tex
\section{State-Aware Representation}
\label{sec:state-aware}

The goal of this step it to obtain a state-aware representation  of WiFi CSI values corresponding to an activity which encodes both the \emph{local} features as well as the \emph{contextual} features of an activity. \bnote{The local features refer to the frequency response of an activity at a particular time-step. In contrast, the contextual features learn the contextual relationship among the local features.} At the end of this step, we obtain an information-rich, lower-dimension vector representation of a CSI stream that captures both local and contextual features of an activity. This state-aware feature is generic, and can be used directly with an off-the-shelf supervised classifier to recognize \emph{seen} activities\footnote{In Section~\ref{sec:eval_seen}, we conduct an experiment to demonstrate its performance.}. In \Sys, we use this representation as an intermediate output, which is used by the later stages of the processing pipeline to enable recognition of both \emph{seen} and \emph{unseen} activities.     



\subsection{The Need for State-Aware Representation} 

The CSI stream is a time series of complex numbers, sampled at a rate of between 200Hz~\cite{virmani2017position} to 2.5KHz~\cite{wang2015understanding}. The raw CSI data are not practical for direct use in the training or in the classification steps, as with high sampling frequency, the dimension of the input vector becomes too large to process efficiently. Moreover, raw CSI signals do not explicitly reflect the intrinsic characteristics of the high-level activity, and generally, they contain artefacts of environmental factors that are not related to the target activity. Hence, like most sensing and inference systems, we map raw data to a lower dimensional vector, commonly known as the \emph{feature vector}, based on the time and frequency domain properties of the signal. 

A wide variety of features have been proposed in the literature for activity recognition from WiFi CSI data. The list includes classic time and frequency domain features such as cepstral coefficients and spectral energy~\cite{wang2014eyes}, as well as recent practices where a layer of a trained deep neural network is considered as the learned features~\cite{yue2018extracting}. However, none of these techniques consider the sequential nature of an activity when converting raw CSI values to feature vectors. 


\begin{figure}[!htb]%
\centering
\subfloat[Two Examples of \emph{`Run'}]{%
\includegraphics[height=.65in]{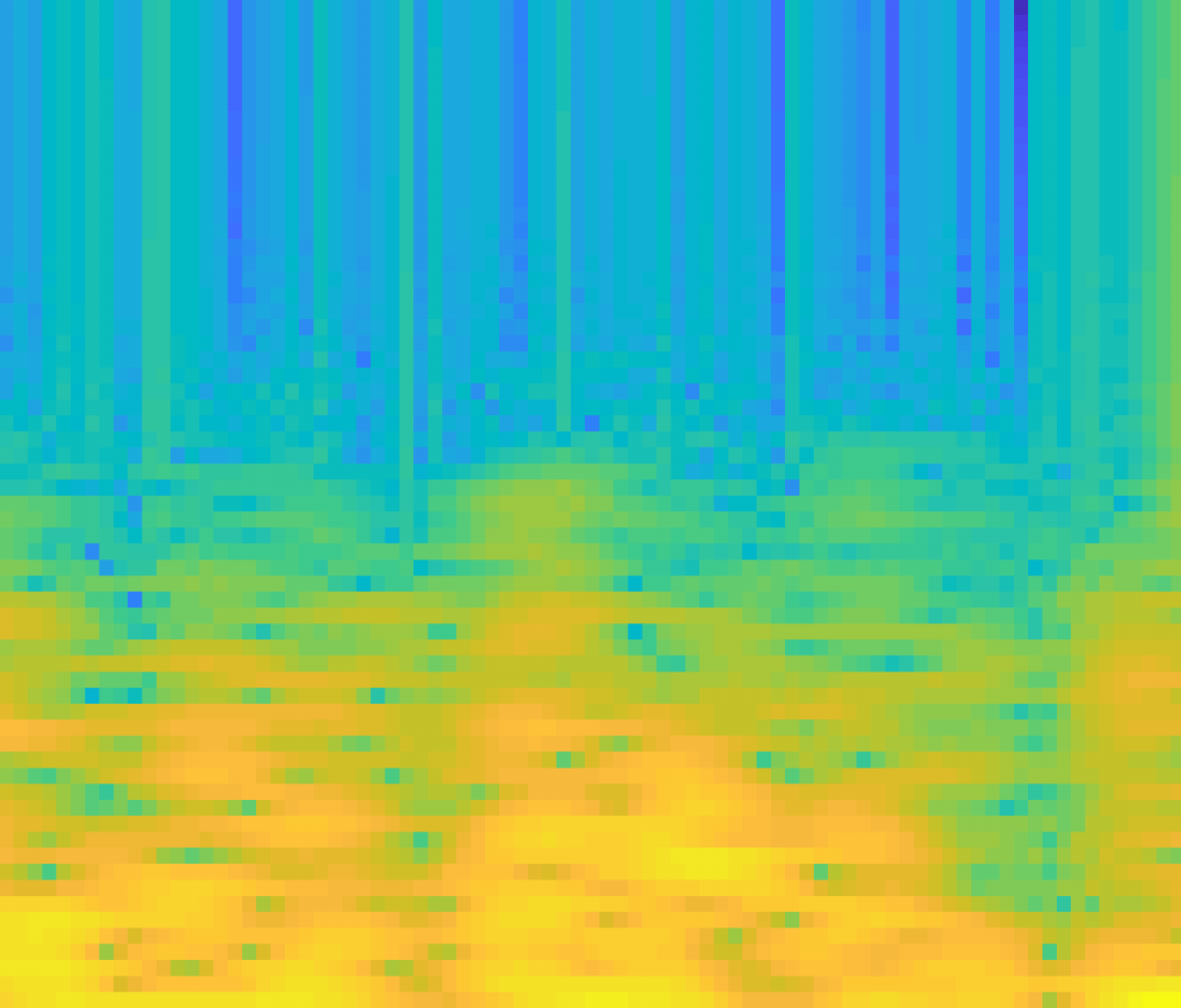}
\includegraphics[height=.65in]{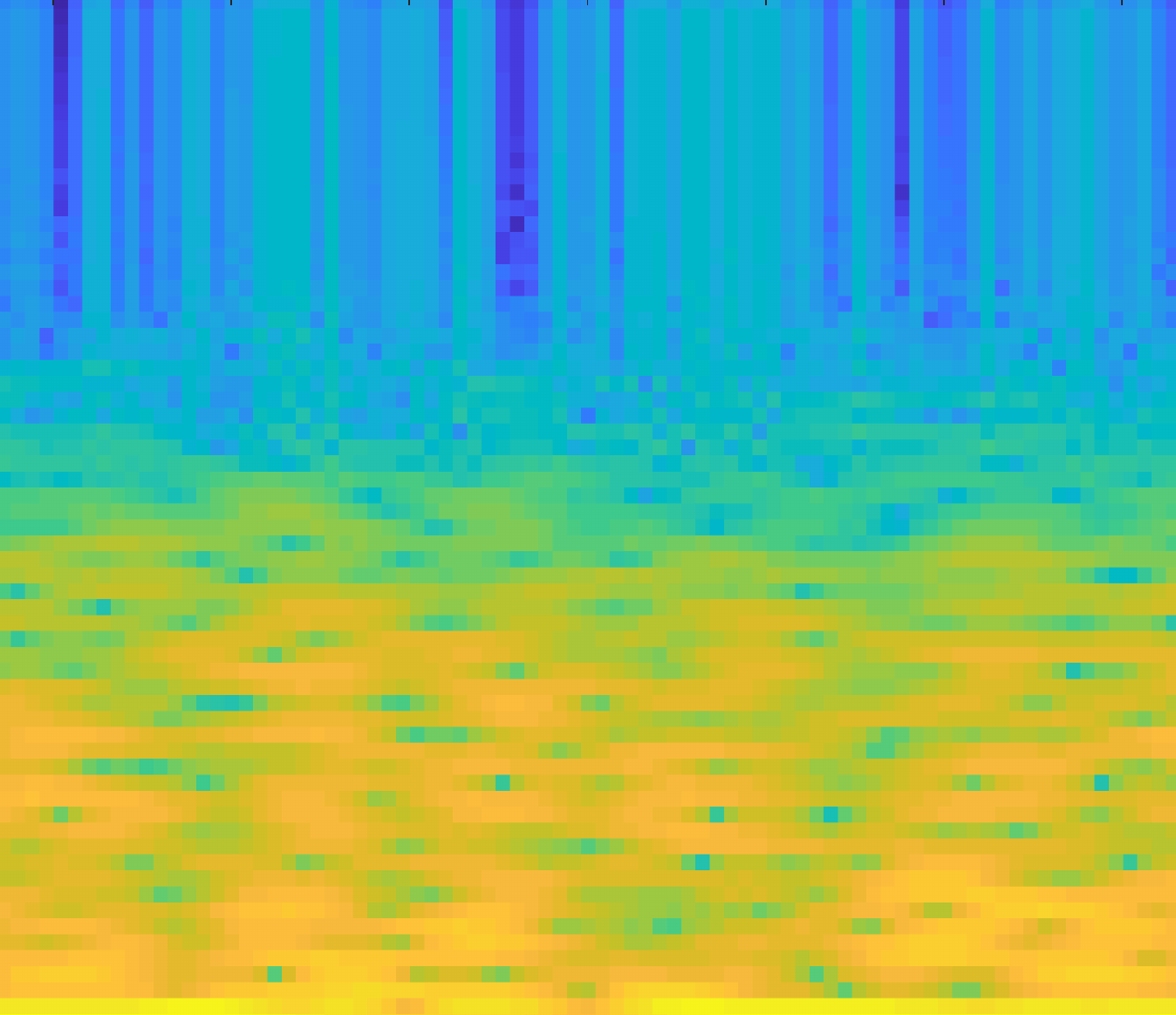}
}%
\enspace
\subfloat[Two Examples of \emph{`Push'}]{%
\includegraphics[height=.65in]{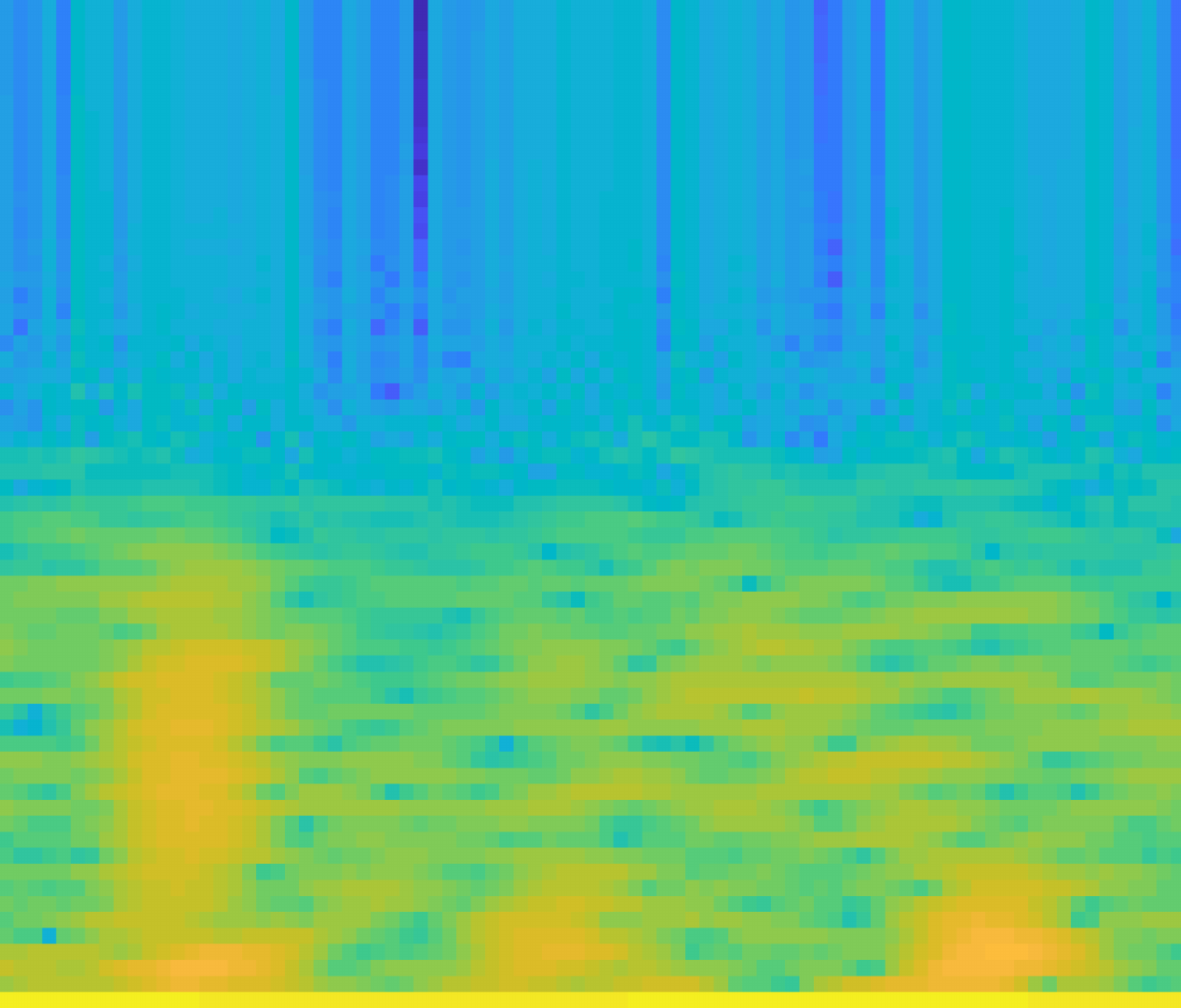}
\includegraphics[height=.65in]{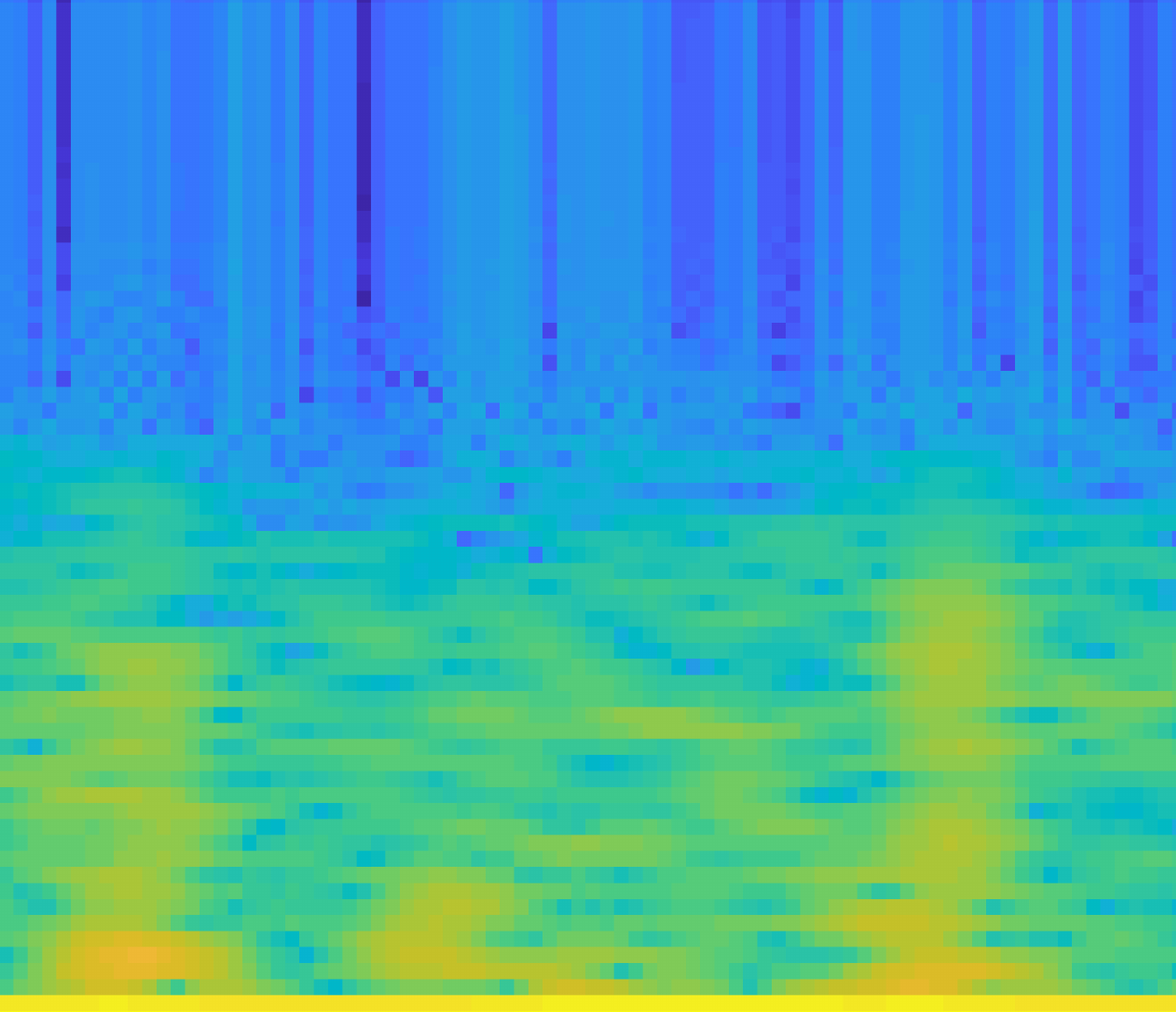}
}%
\caption{Different activities have different state sequences.}
\label{fig:state-seq}
	\vspace{-1.55em}

\end{figure}



Human gestures and activities are time-series of micro-activities that we call \emph{states}. For example, in Figures~\ref{fig:state-seq}, we show four spectrograms of CSI values where the first two correspond to two examples of the \mytt{run} activity, and the last two corresponds to two examples of the \mytt{push} activity. We observe that the pattern how the frequency response changes over the duration of an activity is similar (if not quite the same) for the same activity, and dissimilar for the two different activities. Hence, the key to recognize an activity is to model the sequential change in its frequency spectrum.       


To capture the sequential properties of gestures and activities, ~\cite{wang2015understanding} uses a \emph{Hidden Markov Model} (HMM)~\cite{srinivasan2017preventive} where traditional time-frequency features (\emph{Discrete Wavelet Transform} (DWT)~\cite{wang2015understanding}) are used to represent the states rather than learning representation that embeds the sequential properties of states. A fundamental limitation of this and any other classical HMM-based approaches is that the Markovian assumption, i.e., a state depends only on the previous state, does not hold for most activities. These models do not work when the number of states is large and variable, and an activity has longer term dependencies between the micro-activities that constitute it. 

\begin{figure}[!htb]
    \centering
    \includegraphics[width=0.45\textwidth]{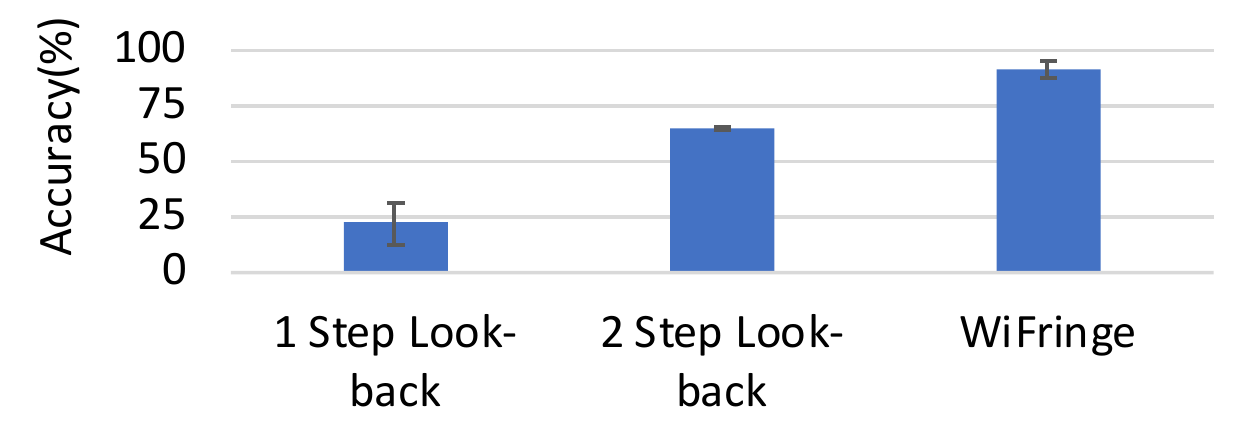}
    \caption{Looking back deeper in time improves modeling accuracy. }
    \label{fig:markov}
	\vspace{-1.0em}

\end{figure}




To demonstrate this, we conduct an experiment to predict the next state of an activity based on the previous states. We use five activities \{\mytt{push, pull, run, walk, eat}\} and divide each sample into five segments that represent the states that constitute an activity. To test the Markovian assumption about the longer term dependencies among the micro-activities, we consider two variants: (a) \emph{One step Look-Back}: a five-state Markov model resulting in a $5 X 5$ dimension transition matrix, $A$, where each entry, $A_{ij}$ represents the transitional probability of going from state $i$ to state $j$, and (b) \emph{Two Step Look-back}: a 25-state Markov model resulting in a $25 X 5$ dimension transition matrix, $A$, where each entry, $A_{ijk}$ represents the transitional probability of going from state $ij$ to state $k$. In Figure~\ref{fig:markov}, we observe that using the one step look-back, we get an average accuracy of $20\%$ for predicting the next state. We observe an improvement of prediction accuracy ($63\%$) when we use the two step look-back variant, which conditions on previous two states to predict the next state. This tells us that a strong Markovian assumption that the current state only depends on previous state does not hold strongly for activity's sequence of states and looking back deeper in time improves the modeling accuracy.  

To overcome the limitations of existing activity modeling techniques, we propose \emph{state-aware} feature representation of CSI streams that captures complex, non-linear dependencies between micro-actions that constitute an action---without requiring a strict Markovian assumption or a predefined, fixed set of states. To achieve this, we employ state-of-the-art deep neural networks that capture both the local (static) as well as contextual (sequential) properties of CSI spectrograms. Our proposed state-aware model is able to predict the next state of an activity with  $90\%$ accuracy(Figure~\ref{fig:markov}), which is $30\%-70\%$ higher than Markovian models. 



\subsection{Rationale Behind Deep Neural Networks}

In \Sys, a \emph{Convolutional Neural Network} (CNN)~\cite{goodfellow2016deep} and a \emph{Recurrent Neural Network} (RNN)~\cite{goodfellow2016deep} are used in tandem to capture the local and the contextual features, respectively.  

\begin{figure}[!htb]
    \centering
    \includegraphics[width=0.45\textwidth]{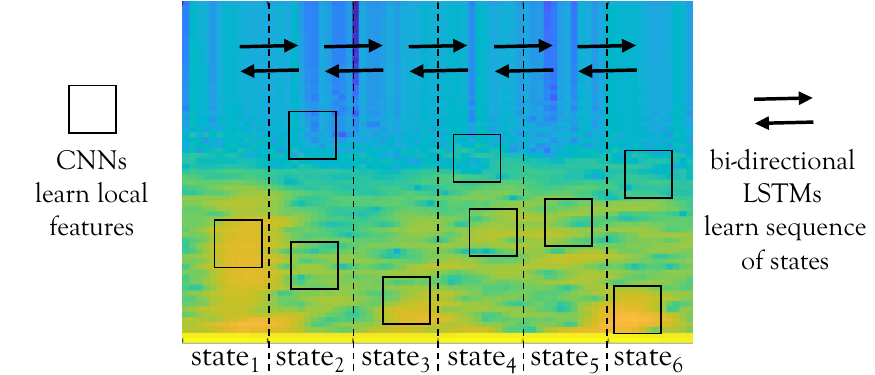}
    \caption{CNNs learn local patterns that characterize each state, whereas RNNs (LSTMs) learn sequence of states.}
    \label{fig:cnn_rnn_justify}
\end{figure}


The CSI spectrogram exhibits rich, informative, and distinguishable patterns for different states within an activity. It contains temporal information, for which, a CNN is the most suitable choice~\cite{goodfellow2016deep}. CNNs contain a hierarchy of filters, where each filter's job is to detect the presence of a particular pattern in a small region on the spectrogram. In Figure~\ref{fig:cnn_rnn_justify}, we use rectangular boxes on the spectrogram which are recognized by the different convolutional filters of the CNN shown on the left.

To model the sequential variation of states within an activity, we use an RNN. We choose an RNN over classical approaches such as a Hidden Markov Model (HMM) since unlike HMMs, RNNs do not require a strong Markovian assumption~\cite{delsole2000fundamental}. With RNNs, a model can learn long-term dependencies among the states. Furthermore, neural networks are in general better at learning complex patterns within the data than shallow models, and RNNs do not need fixed states like an HMM does. In Figure~\ref{fig:cnn_rnn_justify}, we use slices of a spectrogram to denote states and use arrows to illustrate the inherent dependencies between nearby states. 

To construct the RNN, we choose \emph{Long Short Term Memory} (LSTM) as the recurrent component since LSTMs are better at learning long-term dependencies between states~\cite{goodfellow2016deep}, which makes them suitable for capturing relationship between the past states with the recent states. To preserve contextual information from both the future and the past states, we use a bi-directional LSTM model. \bnote{A unidirectional forward LSTM only captures the state transition from past to recent states. But a bi-directional LSTM learns how future states influence past states as well. In many cases, it is important to bring information from the last few states to model an activity and bi-directional LSTMs have this feature.}

\begin{figure}[!htb]
    \centering
    \includegraphics[width=0.4\textwidth]{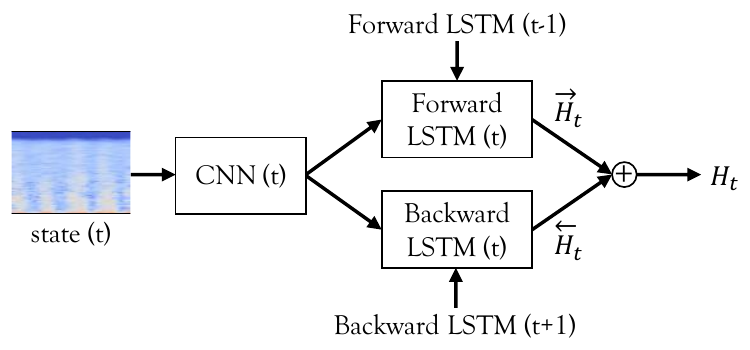}
    \caption{Network architecture for state-aware representation.}
    \label{fig:conv_lstm}
	\vspace{-1.55em}

\end{figure}

\subsection{Detailed Algorithmic Steps} 

Figure~\ref{fig:conv_lstm} shows the integrated neural network architecture that takes CSI spectrogram as the input and produces the state-aware vector representation through a sequence of processing steps. 

$\bullet$ \emph{CSI Processing:} For a given CSI stream, $X$ that corresponds to an instance of an activity, we divide the CSI values into $n$ \bnote{equal} segments $\{X_1, X_2, \dots, X_n\}$. Here, $n$ is empirically determined, we used $n=5$ in all our experiments. For each segment $X_i$, we take the spectrogram~\cite{lu2009deconvolutive} to obtain $S_i$ as follows:
\[
\mathrm{S_i = STFT(X_i)} 
\]
where, STFT(.) denotes \textit{Short-Time Fourier Transform}~\cite{lu2009deconvolutive}, which estimates the short-term, time-localized frequency content of $X_i$.

$\bullet$ \emph{CNN Processing:} We use a three-layer CNN, $G_{\theta}$, where $\theta$ are the parameters of the network, which takes $S_i$ as the input and produces a $1,000$ dimension vector, $L_i$ that represents the local features of the input spectrogram:
\[
\mathrm{L_i = G_\theta(S_i)} 
\]


Each layer of the CNN extracts a feature map from the input data. As we go deeper into network, the deeper-layer filters extract more information. We empirically determine that three layers of convolutional operations extract adequate information for the later steps of \Sys. The filters have a size of $3 X 3$ and we increase the depth from $16$ to $64$ channels for the three layers.  We use \emph{Rectified Linear Units} (ReLu), $\sigma(a) = \max\{a, 0\}$ as the non-linear activation function for their robustness against the vanishing gradient problem~\cite{hochreiter1998vanishing}. The output of the third convolutional layer is fed into two \emph{dense} layers~\cite{goodfellow2016deep} with a ReLu activation layer in-between them. We keep $1,000$ neurons in the last dense layer, which is the dimension of the local feature vector. 

$\bullet$ \emph{RNN Processing:} For a total of $n$ segments, we obtain $n$ local feature maps $G_{\theta}(S_i)$ from the CNN. Each state's local feature is fed into a bi-directional LSTM (bi-LSTM) to model the contextual property of the states of an activity. The bi-LSTM network has two unidirectional LSTMs, i.e., a \emph{forward} and a \emph{backward} LSTM. For the forward LSTM, each hidden state $\overrightarrow{H_i}$ depends on the previous state $H_{i-1}$ and the input $S_i$. On the other hand, for the backward LSTM, each hidden state $\overleftarrow{H_i}$ depends on the future state $H_{i+1}$ and the input  $S_i$. By using a bi-directional LSTM, we obtain a representation of an activity that is aware of the transition of states which yields a strong model of the activities. We use 500 neurons in both the forward and the backward LSTMs. Each LSTM has a dropout layer to avoid overfitting.

$\bullet$ \emph{Representation:} The final  hidden representation of the bidirectional LSTM is the concatenation of $\overrightarrow{H_i}$  and $\overleftarrow{H_i}$.

%% file: tex/06B_CMP.tex
\section{Cross-modal Projections}
\label{sec:CMP}

In the previous section, we proposed a feature learning algorithm that learns to map CSI stream segments onto a feature space that  captures the contextual information of sequential states of each activity. This gives us a robust activity representation, which by itself can directly be used to train robust activity classifiers. However, that representation can not be used to perform zero-shot learning as it does not borrow knowledge from an external source. In this section, we describe the cross-modal projection step of \Sys which brings external knowledge from the text domain to enable classification of unseen activities. 

\subsection{The Need for Cross-Modal Projections}

\gnote{Existing CSI-based human activity recognition systems can recognize only a fixed set of activities for which the system has been trained for. When an unseen activity is presented to these systems, the best they can do is to find the class that is closest to the given activity. Due to their reliance on a single source of knowledge (i.e., only WiFi CSI-based training data), these systems by design cannot generate a new class label that describes the unseen example. By having additional sources (or \emph{modes}) of knowledge, e.g, the relationship among words and their attributes in the English text, we can enhance the ability of these systems to recognize previously unseen activities, and to generate class labels that come from the additional knowledge sources (i.e., the text domain).}         

The secret recipe behind \Sys's ability to classify unseen activities is the \emph{cross modal projection}. Through this step, we blur the difference between an activity's CSI stream and its corresponding English word embedding and attributes, and make them (almost) equal in their feature representations. In other words, if an English word (say, \mytt{run}) has a known vector representation (e.g., obtained using Google's Word-2-Vec~\cite{mikolov2013efficient} on a large dataset like Wikipedia) and a vector of attributes of the word (e.g., \emph{movement of legs} and \emph{motion of hands}, which is obtained from~\cite{zellers2017zero}), the goal of the cross-modal projection is to generate the exact same vectors when \Sys is presented with a CSI stream of that word (i.e., CSI of running). 

\begin{figure}[!thb]
    \centering
    \includegraphics[width=0.48\textwidth]{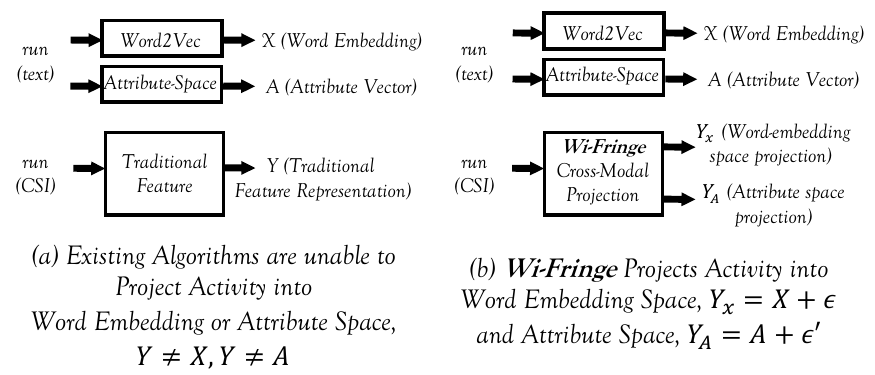}
    \caption{Cross-modal projections map a CSI stream to its corresponding label in the word and attribute embedding spaces.}
    \label{fig:cm-pipeline}
    \vspace{-1.0em}
\end{figure}


Figure~\ref{fig:cm-pipeline} illustrates cross-modal projection using \mytt{run} as the example activity, whose word embedding and attribute vectors are $X$ and $A$, respectively. In \Sys, the activity representation of \mytt{run} is passed through two cross-modal projection steps to obtain $Y_x$ and $Y_A$, corresponding to the word and the attribute spaces, respectively, and it is made sure that $Y_x \approx X$ and $Y_A \approx A$. In traditional feature representation, however, the projection is different (i.e., $Y \neq$ X and $Y \neq$ A), and hence, unlike \Sys, traditional features cannot guess the word label of an unseen activity.

\subsection{Rationale Behind Multiple Projections}

The benefits of cross-modal projection from CSI to two latent spaces, i.e., word and activity-attribute spaces, are as follows: 

    $\bullet$ \emph{Word Embedding Space:} There are over 150 thousand English words for which researchers in the natural language processing field have created semantically aware vector representations, called the \emph{Word Embedding}. Such an embedding preserves the contextual relationship among words and puts two words that are similar in meaning or are often used in the same context closer in the representation space. By projecting the activity representation to the word embedding space, \Sys is able to generate meaningful and context-aware representation of any CSI stream, irrespective of whether or not it has seen CSI of the same class before.

    $\bullet$  \emph{Activity-Attribute Space:} Human activities and bodily gestures comprise of movements by different body parts, i.e., arms, legs, head, and torso. Activities also involve external objects, e.g., an \mytt{eating} activity may involve the use of spoons and knives. These \emph{attributes} create nuances in different RF-based activity feature representations. Projecting CSI onto the activity-attribute space embeds this information into the projection, as CSI implicitly gets affected by moving different sorts of objects due to their reflections. Using this embedded contextual and attribute information from CSI, \Sys recognizes activities without any training examples.

\Sys uses both word embedding and activity-attribute spaces for cross-modal projections to combine the predictive power of both. Combining these two help each other in the final prediction step. For example, in the word embedding space (50 dimensional W2Vec), \mytt{run} is surprisingly closer to \mytt{pull} than \mytt{walk}, since they appear more with similar neighbouring words. However, \mytt{run} and \mytt{walk} are more similar in terms of activity properties. In the attribute space, however, \mytt{run} and \mytt{walk} are closer to each other. In such scenarios, where relying only on the contextual semantic information fails to figure the correct relationship between activities, the attribute information helps correct the misprediction. 

Likewise, in many cases, the attributes extracted for two activities are very similar. For example, \mytt{run} and \mytt{walk} have the exact same attribute (binary) vectors when the attributes are constructed from online dictionaries~\cite{zellers2017zero}. While we desire their vectors to be closer in the attribute space for these two activities, we do not want them to be exact. This is because if they are the same, the RF projections for these two activities will also be the same. Thus, \Sys will not be able to distinguish between \mytt{walk} and \mytt{run} if we only rely upon the attributes. In such scenarios, the word embedding helps correct the problem and \Sys is able to separate the activities in the joint feature space.  

\subsection{Detailed Algorithmic Steps}



The goal of cross-modal projection is to map state-aware representation of WiFi CSI streams (Section~\ref{sec:state-aware}) to two latent spaces, i.e., the word embedding space and the activity-attribute space. The difficulty in this step is that although the class label corresponding to an activity has a fixed word embedding and a fixed attribute-vector, the CSI values corresponding to the same activity vary due to the diversity in human motion and physique. This makes the mapping from the CSI domain to the word-embedding and activity-attribute spaces a non-linear projection problem. In \Sys, we solve this problem by using a neural network to learn the projection from CSI to the latent spaces. 

\begin{figure}[!htb]
    \centering
    \includegraphics[width=0.45\textwidth]{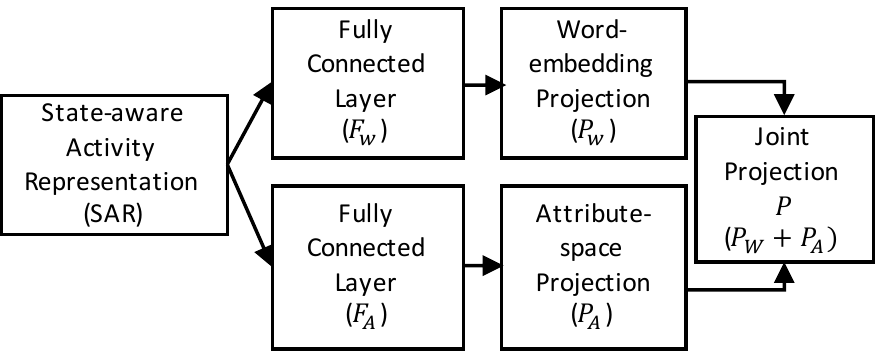}
    \caption{State Aware Representation is projected into both Word-embedding and Attribute space which are then aggregated for a joint projection.}
    \label{fig:joint-cross}
     \vspace{-.75em}

\end{figure}

The projection operation is illustrated by Figure~\ref{fig:joint-cross}. The state-aware representation, i.e., the output of the LSTM from Figure~\ref{fig:conv_lstm}, is fed to two neural networks having fully connected layers. The first network projects the state-aware representation onto the activity attribute space, and the second network projects the representation onto the word embedding space. We refer to the last layer of the neural networks that perform attribute space projection and word embedding projection as $F_A$ and $F_W$, respectively.

$\bullet$ \textit{Projecting onto Activity-Attribute Space.} From the attribute database provided by~\cite{zellers2017zero}, we obtain a set of binary attributes associated with each activity. Each activity, $a_i$ is represented as an $m$-dimension vector, $d_i \in \{1, 0\}^m$, where $m$ is the total number of attributes used to define an activity. Each element $d^{(k)}_i$ is a binary indicator of whether the $k^{th}$ attribute is true or false for that activity.  



\begin{equation}
    d^{(k)}_{i}=
    \begin{cases}
      1, & \text{if attribute $k$ is true for activity $a_i$.}  \\
      0, & \text{otherwise}
    \end{cases}
\end{equation}

To get the likelihood of a CSI stream being predicted as an activity $a_i$, we project  $F_A$ to the attribute space. We take the dot product of the attribute vector $d_i$ and $F_A$. The dot product demonstrates the similarity between the projection $F_A$ with attribute vector 
$d_i$. For a CSI stream from activity $a_i$, our model's target is to increase the similarity between $d_i$ and $F_A$. The similarity which denotes the likelihood of $F_A$'s probability of belonging to activity $a_i$ in attribute space is calculated as a dot product:

\begin{equation}
    P^i_A = d_i \cdot F_A
    \label{eq:proj_a}
\end{equation}




$\bullet$ \textit{Projecting onto Word-Embedding Space.} For an activity, $a_i$ whose word embedding is $w_i$, we want to project the CSI-based state-aware representation as close as possible to $w_i$.  \bnote{Therefore, for a CSI stream of an activity $a_i$,  our target is to project  $F_W$  to be close to $w_i$ in vector space
} This results in a higher value of dot product between $F_W$  and $w_i$. Therefore, the similarity in word embedding space is defined as following: 
\begin{equation}
    P^i_W = w_i \cdot F_W
    \label{eq:proj_w}
\end{equation}

$\bullet$ \textit{Projecting onto the Joint Space.} To obtain a joint projection on both the attribute and the word embedding space, we employ an ensemble approach to combine the two projections from Equations~(\ref{eq:proj_a}) and~(\ref{eq:proj_w}) as follows: 
\begin{equation}
    P_i =P^i_A +P^i_W
\end{equation}
where, $P_i$ carries the confidence of a CSI segment's probability of belonging to class $a_i$. For |a| number of activities, given $F_A$ and $F_W$ extracted using state-aware representation for a CSI segment $X$ as described in Section~\ref{sec:state-aware}, the probability of $X\in a_i$ is calculated using the following softmax operation:
\begin{equation}
\label{eqn:softmax}
    p(a_i|F_A,F_W)=\frac{e^{P_i}}{\sum^{|a|}_{j=1} e^{P_j}}
\end{equation}


%% file: tex/06C_classify.tex
\section{Two Stage Classifier} 
\label{sec:classify}



\Sys employs a two-stage classifier to infer the most likely activity type for an input CSI stream segment. The first-stage classifier determines whether the input CSI stream segment belongs to a \emph{seen} or an \emph{unseen} class. The second-stage classifier makes the final determination of the most probable activity type for the input CSI stream segment. 



\subsection{The Need for Two-Stage Classifier}

\begin{figure}[!htb]
    \centering
    \includegraphics[width=0.45\textwidth]{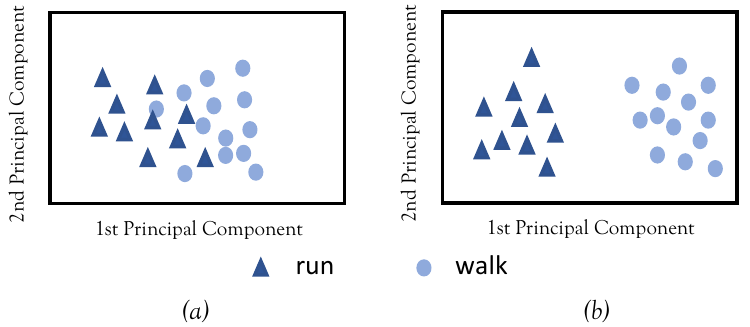}
    \caption{(a) State-aware activity representation projects samples of unseen class \emph{run} close to samples of seen class \emph{walk}  (b) Spectrogram level feature  projects the samples of unseen class \emph{run} far from the samples of seen class \emph{walk}.}
    \label{fig:two-step-classifier}
 	\vspace{-1.15em}

\end{figure}

Neural networks tend to memorize patterns in data from training. Thus, even for an unseen category, the network tries to project an input close to one of the seen classes in the state-aware representation space. While this serves our purpose of classifying an unseen activity, it affects the classification performance of a generalized system where the system may encounter examples from both seen and unseen classes. Since the unseen classes are projected too close to the seen classes, they will be classified as one of those seen classes. In Figure~\ref{fig:two-step-classifier}(a), we plot the first two principal components of our proposes state-aware activity representation for two activities: \mytt{run} and \mytt{walk}. Here, the state-aware representation is trained with samples from \mytt{walk}. We see that for these two activities, the projections are very close to each other with some overlaps. This results in errors in deciding the samples from \mytt{run} (an unseen category) and they get classified as \mytt{walk}.

To overcome the problem posed by neural network based state-aware representation in distinguishing seen vs. unseen categories, we employ a classifier which is based on the signal characteristics such as the time-frequency representation. To be more precise, while some activities such as \mytt{run} and \mytt{walk} have similarity in their attributes, they still have distinguishable signal characteristics that are embedded in WiFi CSI. Past works~\cite{wang2015understanding, wang2014eyes} have proven the distinguishable power of traditional signal-level features such as Short Time Fourier Transform (STFT) and Discrete Wavelet Transform(DWT). We use STFT to detect if a sample is from a seen or an unseen category. STFT gives us the changes in frequency components of the signal along the time axis.  

In Figure~\ref{fig:two-step-classifier}(b), we  plot the first two principal components of the STFT for the samples \mytt{run} and \mytt{walk}.  We see that they are well separated in the STFT representation, which allows us to determine if an example is from a seen or an unseen category. 

\subsection{Detailed Algorithmic Steps}

The two steps of the algorithm are as follows:

$\bullet$ \textit{Seen vs. Unseen Detection.} We devise a simple threshold-based decision algorithm to determine whether an input CSI stream segment belongs to an \emph{unseen} class. We use K-means~\cite{krishna1999genetic} clustering algorithm to cluster the STFT of the training samples of the seen category classes. This gives us $K$ cluster centers, ${C_1,C_2,\dots,C_K}$ for $K$ seen classes. For an input CSI stream segment, $u$, it belongs to an \emph{unseen} class if the following condition is true:
\begin{equation}
    \min_{s \in S} \| C_s - STFT(u) \| > \Omega
\end{equation}
where, $S$ is the set of \emph{seen} classes and $\Omega$ is an empirically determined threshold that maximizes the accuracy of \emph{seen} vs. \emph{unseen} class detection, and $\|.\|$ is the Euclidean norm. When this condition is false, $u$ belongs to a \emph{seen} class.

$\bullet$ \textit{Classification.} If the CSI segment is recognized as from a \emph{seen} category, only the labels from seen category are considered and the class label is obtained by applying the following equation:
\begin{equation}
    \argminB_{a_i} p(a_i|F_A, F_W)
    \label{eq:clabel}
\end{equation}
On the other hand, if the CSI segment is recognized as from an \emph{unseen} category, we exclude all the labels from the \emph{seen} category and only the labels from the \emph{unseen} category are considered and the class label is obtained by applying Equation~\ref{eq:clabel}.  


%% file: tex/07_implementation.tex
\section{Experimental Setup}
\label{sec:impl}

\begin{figure}[!htb]%
\centering
	\includegraphics[width = 0.45\textwidth]{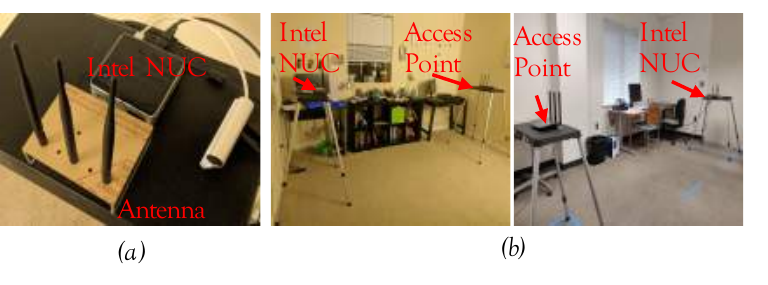}
	\caption{(a) Intel Nuc with Antennas. (b) Experimental Setup.}
	\label{fig:exp-setup}
	\vspace{-1.5em}
\end{figure}

\subsection{WiFi CSI Collection} 

We implement \Sys on an Intel NUC~\cite{nuc-pc} mini PC. The mini PC is interfaced with an Intel NIC 5300~\cite{nic-card} and three omni-directional antennas with 6 dBi gain. To extract CSI values from the NIC card, we use the tool developed by Halperin et. al~\cite{halperin2011tool}. This tool provides a modified firmware for NIC 5300 and an open source Linux driver with user space tools to collect CSI information of the received packets. We use a Netlink router as an access point (AP). The mini PC pings the AP periodically at a certain interval and the CSI is extracted from the packet that is received by the mini PC as an echo from the AP. The sampling rate of ping is set at 500 sample per second. The PC and AP are placed 12 feet apart from each other. Experiments are done in a student apartment and an office space.  Both rooms have typical furniture, e.g., two tables, multiple chairs, and cabinets. Activities are performed by four volunteers (1 female and 3 males). Figure~\ref{fig:exp-setup}(a) shows the mini PC Intel NUC connected to three antennas which we place on a stand during the experiments. Figure~\ref{fig:exp-setup}(b) shows the the two rooms with our experimental setup.

\subsection{Noise Reduction}

CSI data extracted from the NIC card suffer from dominant noise, which makes gesture recognition  difficult. To remove noise, we follow~\cite{wang2015understanding}'s PCA-based denoising technique. Since the Channel Frequency Response (CFR) for different subcarriers is a linear combinations of the same set of waveforms with different initial phases, when a gesture is performed, the changes of CFR value in different subcarriers are correlated. To extract the changes in CFR values caused by the movement in different subcarriers, we apply PCA on the CSI stream collected from 30 subcarriers for each TX-RX antenna pair. The ordered principal components give us the most variances experienced across all sub-carriers. We discard the first principal component stream as it contains dominant effect of noise. The second and the third components contain clear effect of gestures and less effect of noise. We choose the third stream as it has shown better noise immunity. Although the selected stream is less prone to noise, it still contains some effect of unwanted high-frequency noise. Hence, the stream is passed through a Butterworth filter~\cite{selesnick1998generalized} to remove static noise. We set a cut off frequency of 50Hz as human activity is usually  below 50Hz in frequency. Thus, we obtain a de-noised CSi stream from all sub-carriers.
 



\subsection{Gesture Segmentation}

In order to detect the start and the end of a gesture, we instruct the volunteers to perform gestures with a brief pause between them. As the peaks in a CSI stream depends on the initial position of the user, which is variable for different gestures, we cannot segment gestures by directly applying a threshold on the CSI values. Instead, we follow~\cite{virmani2017position} and take the first order difference of the stream which is stable during the pause and fluctuates during gestures. We apply a threshold-based peak detection technique on the first order different to detect gestures. The segments on the first order difference that are above a predefined threshold are detected as gestures. We refer~\cite{virmani2017position} to the readers for more details on pre-processing of CSI stream.


 \subsection{Data Augmentation} 
 
 To avoid overfitting and to make our model robust, we use a data augmentation technique to increase the user contributed data size by about 20 times. First,  we apply a geometric transformation to the time-frequency representation of the raw CSI signals. Specifically, we use translation to shift the spectrogram along the time axis. We exclude shifting along the frequency axis, as it does not reflect any physical world phenomenon. Shifting along the time axis reflects the effect of recording the same gesture at different timestamps. Second, We superimpose simulated noise on the signal to augment the dataset further. Each signal is superimposed with a Gaussian white noise~\cite{islam2017soundsifter}. The noise model is generated offline. Injecting noise captures environmental effect and makes the model resilient to noisy environment.

\subsection{Empirical Dataset} 

In this section, we describe our dataset collection method. \Sys leverages knowledge from the text domain using the activity names to classify activities without training examples. For that reason, we need \emph{named} activities that have semantic representation in English language. For example, \Sys is able to classify \emph{run} if it has seen training examples of \emph{walk} based on the semantic relationship between these two words. However, \Sys won't be able to classify activities without having semantic meaning in English or any other language.

Based on our study of named activities from~\cite{anguita2012human,anguita2013public,venkatnarayan2018multi}, we collect $20$ most common named activities for our empirical evaluation. To the best of our knowledge, this is the largest dataset in terms of the number of classes considered in WiFi-based activity recognition systems. Existing literature have used 6--8 types of gestures for supervised activity recognition from WiFi~\cite{virmani2017position, venkatnarayan2018multi, wang2014eyes, wang2015understanding}. Our dataset contains activities collected from four volunteers in two different rooms with different orientations and furniture. Our dataset is diverse and it stresses out the algorithmic components of \Sys. In Table~\ref{tab:act-list}, we provide the list of the $20$ activities clustered with major attributes. On average, each class have $100$ samples.

\begin{table}[!htb]
 \centering
\resizebox{.38\textwidth}{!}{
\begin{tabular}{r|l}

\textbf{Category}            & \textbf{Activities}  \\ \hline
Freehand Gestures   & Point, Raise, Rub, Scratch, Shake,\\ 
           & Toss, Circle, Arc.\\ 
Object-Human Interactions    & Drink, Eat, Push, Pull.\\ 
Upper/Lower-Body Gestures & Sit, Stand, Bow, Duck, Kick.\\ 
Mobility    & Jump, Walk, Run.\\ 
\end{tabular}
}
\caption{Twenty categories of activities are used in the experiments. Each category has 100 examples on average.}
\label{tab:act-list}
\vspace{-1.55em}

\end{table}


 

%% file: tex/08_evaluation.tex
\section{Experimental Results}
\label{sec:eval}
\subsection{Accuracy of Unseen Class Detection} 

In this experiment, we report the accuracy of \Sys for \emph{unseen} classes. These are the classes for which \Sys did not have any training examples during training phase. As state-of-the-art systems are not capable of recognizing activities without prior training examples, we are unable to compare them with our solution (i.e., any existing algorithm will have an accuracy of random guess at best). Hence, we report \Sys's performance in this section by varying the number of unseen classes and compare it with two variants of our algorithm: (1) projecting State-Aware Representation (SAR) onto only word embedding (W2Vec) space, and (2) projecting State-Aware Representation (SAR) on only activity-attribute space. This comparison shows the performance boost due to joint space projection. Note that \Sys is able to correctly label an unseen activity only if it has learned the representation of classes that are semantically related to it in terms of movement of body parts and word-level embedding. Therefore, while selecting \emph{unseen} classes, we keep at least one class from the \emph{seen} classes which is close to it in word embedding and attribute space. For example, we select \emph{push} as an unseen class and keep \emph{pull} as a seen class which are semantically related. A \emph{push} is similar to a \emph{pull} in motion of the body parts, e.g., both require movement of hand(s) but legs remain unmoved.  Furthermore, the two English words \emph{push} and \emph{pull} are closer in the word embedding space as they appear in similar contexts.

\begin{figure}[!htb]
	\centering
	\includegraphics[width = 0.45\textwidth]{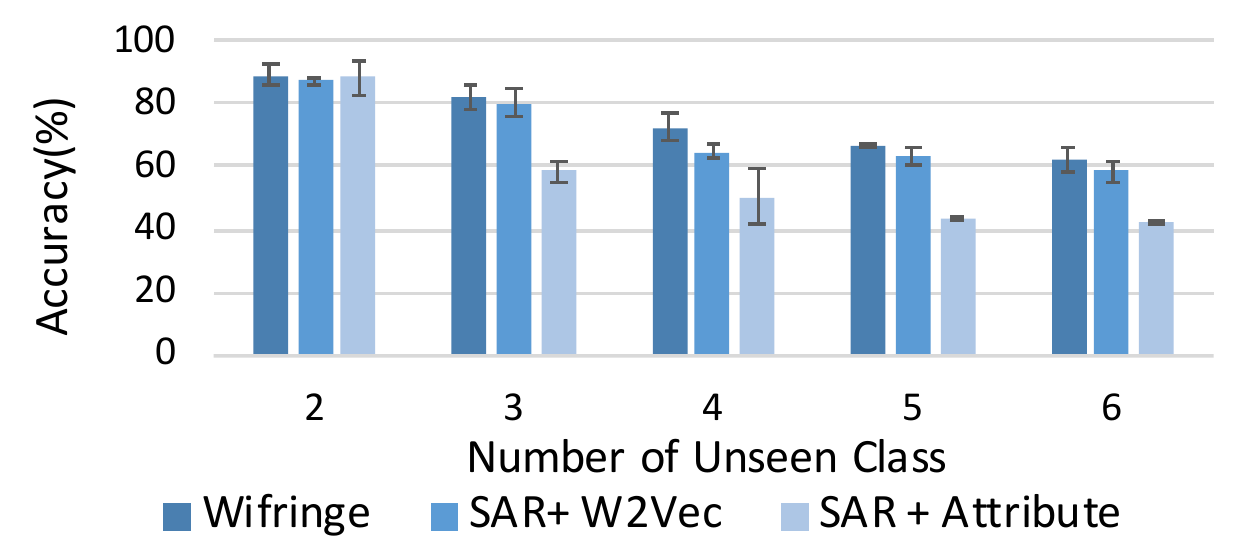}
	\caption{\Sys's accuracy for unseen activity classification.}
	\label{fig:zsl}
	\vspace{-1.05em}
\end{figure}

In Figure~\ref{fig:zsl} we report the accuracy of \Sys in recognizing the unseen classes of activities when 2--6 types of activities are from unseen classes. We evaluate with different combinations of \emph{seen} and \emph{unseen} activities and present the mean accuracy and variance in the plot. In Figure~\ref{fig:zsl}, we see that for two unseen classes, we achieve a classification accuracy of around $90\%$. With only word embedding and attribute space projection, the accuracy is $87\%$ and $88\%$, respectively. For three unseen activities, we get an accuracy near $83\%$ with \Sys. With only word embedding projection the accuracy is around $80\%$, but with attribute space projection the accuracy drops to $60\%$. This is due to the similarity of activities in attribute space, which results in very similar attribute vectors. Therefore, projecting only on attribute space makes the classification harder. As the number of unseen classes increase to 4, 5, and 6, the accuracy becomes to $73\%$ , $67\%$ and and $62\%$, respectively. In all the cases, joint space projection boosts the performance in comparison with single space projection. As the number of unseen classes increase, the problem becomes harder since the model has to differentiate between more classes without training data. We report up to six unseen classes in the plot, however, for seven unseen classes, our accuracy is around $53\%$. With random selection, the accuracy for seven unseen class is $14.28\%$, so \Sys is still better by almost $40\%$.

\subsection{Accuracy of Seen Class Detection} 
\label{sec:eval_seen}

In this experiment, we evaluate \Sys's \emph{seen} class detection performance by keeping \rnote{all} the classes in seen category. We compare \Sys with other baseline classification algorithms. Recent works~\cite{yue2018extracting,jiang2018towards} have demonstrated the efficiency of using Convolutional Neural Network(CNN)-based classifiers for WiFI based activity recognition. Following these works, we compare \Sys with a CNN classifier optimized for our dataset with five convolutional layers along with batch normalization and dropout layers. As mentioned earlier, state-aware representation (SAR) by itself can be used as a feature for supervised classification. Therefore, we also report the performance of state-aware representation (SAR) integrated with a softmax layer. In addition, we also show the performance of projecting state-aware representation only to word embedding space and attribute space. We use five-fold cross-validation by randomly selecting training and testing examples each time. We also report the classification performance of a shallow classifier with a traditional handcrafted feature (i.e., STFT). We report the mean and variance of classification accuracy.
\begin{figure}[!htb]
	\centering
	\includegraphics[width = 0.45\textwidth]{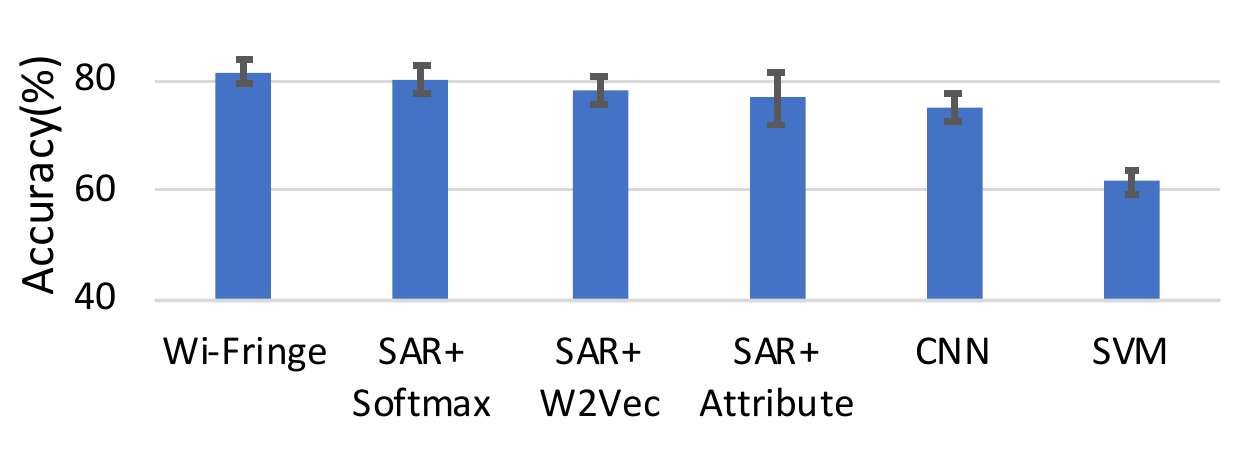}
	\caption{\Sys's accuracy is higher than baseline algorithms in seen class detection.}
	\label{fig:osl}
	\vspace{-.60em}
\end{figure}

In Figure~\ref{fig:osl}, we find that \Sys achieves a mean accuracy of $82\%$. On the other hand, state-aware representation (SAR) along with softmax layer is able to achieve  around $80\%$ mean accuracy. The performance boost of \Sys is due to the fact that from joint space projection, our model is able to classify activities by integrating knowledge from both word embedding and attribute domain. Projecting state-aware representation onto only word embedding and attribute space yields accuracy of $78\%$ and $76\%$, respectively. With CNN, we have an accuracy of $74\%$. Therefore, the proposed state-aware Representation improves the accuracy for seen class detection by $8\%$ than CNN model. This improvement is achieved since the state-aware representation learns the temporal relationship among the states in an activity. With an SVM, we see the accuracy is around $62\%$. Therefore, it is evident that \Sys is able to achieve better accuracy than other classifiers in seen class detection. Note that \Sys's architecture is modular and generic. We can also integrate other representation to \Sys if necessary.

\subsection{Accuracy of Seen vs Unseen Class Detection} 

\begin{figure}[!htb]
	\centering
	\includegraphics[width = 0.45\textwidth]{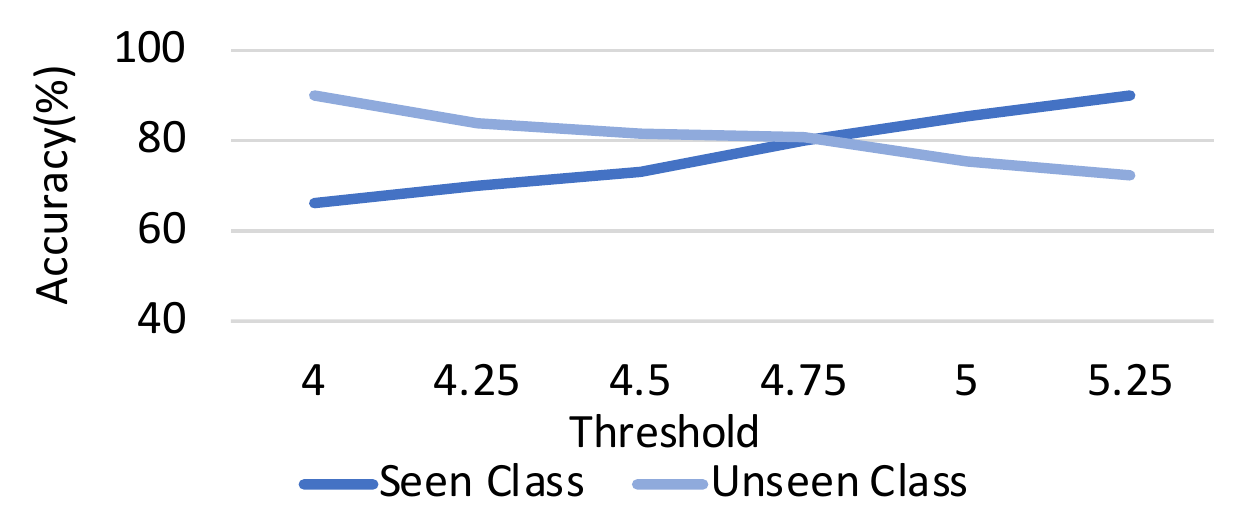}
	\caption{The accuracy of \emph{seen} and \emph{unseen} class detection depends on the threshold  $\Omega$'s value. }
	\label{fig:threshold}
	\vspace{-1.95em}
\end{figure}

In this section, we present the accuracy of our threshold based Seen vs. Unseen detection's performance. The accuracy is threshold dependent. In Figure~\ref{fig:threshold}, we plot the accuracy for $\Omega \in [4.0-5.25]$. We observe that setting a high threshold fails to detect many class as \emph{unseen} and the accuracy drops for the \emph{unseen} classes. With high threshold, the unseen class samples have to be very far apart from any cluster center of the seen class clusters. On the other hand, setting the threshold too small leads to poor results for \emph{seen} classes as it determines majority CSI stream sample as \emph{unseen}. Hence, there is a trade-off between the \emph{seen} and \emph{unseen} class detection accuracy. The optimum threshold is $4.75$, for which, the classification accuracy of the \emph{seen} and \emph{unseen} classes are around $80\%$.

\subsection{End to End Evaluation}

 \begin{figure*}[!thb]
    \centering
    \subfloat[]{{\includegraphics[width=.33\textwidth]{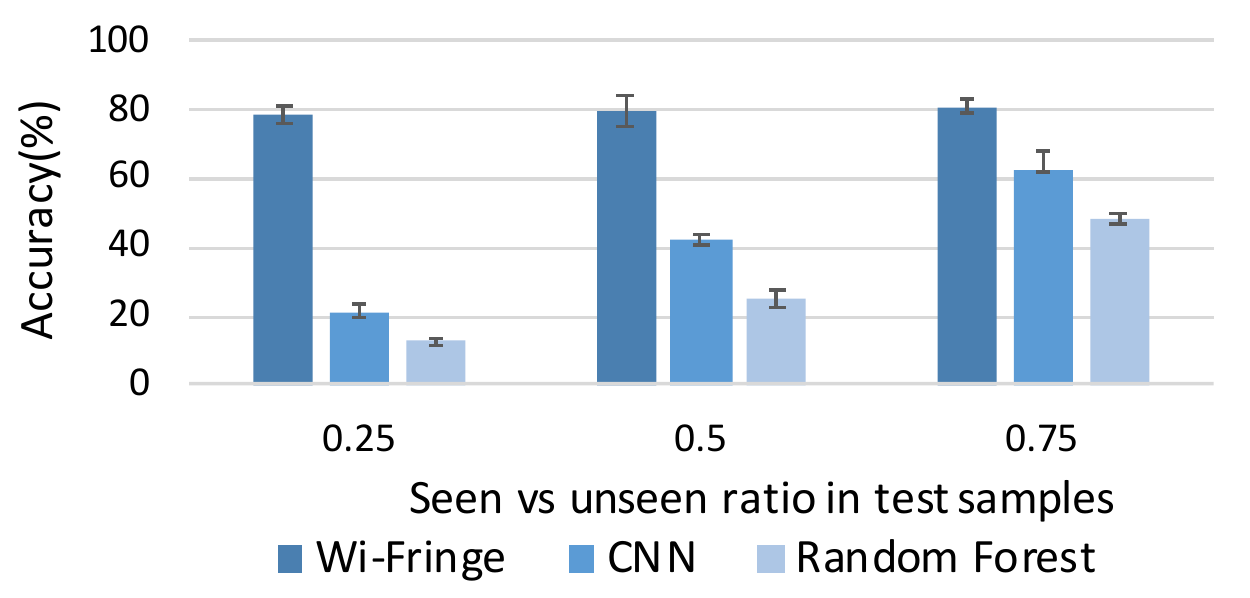}}}%
    \subfloat[]{{\includegraphics[width=.32\textwidth]{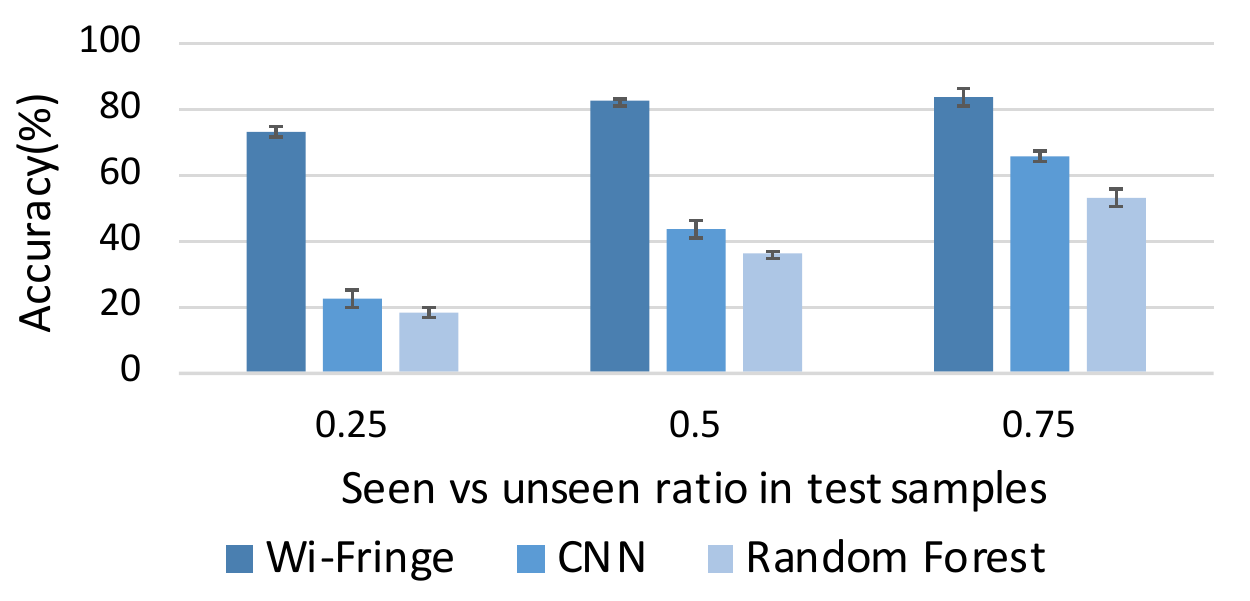} }}%
    \subfloat[]{{\includegraphics[width=.33\textwidth]{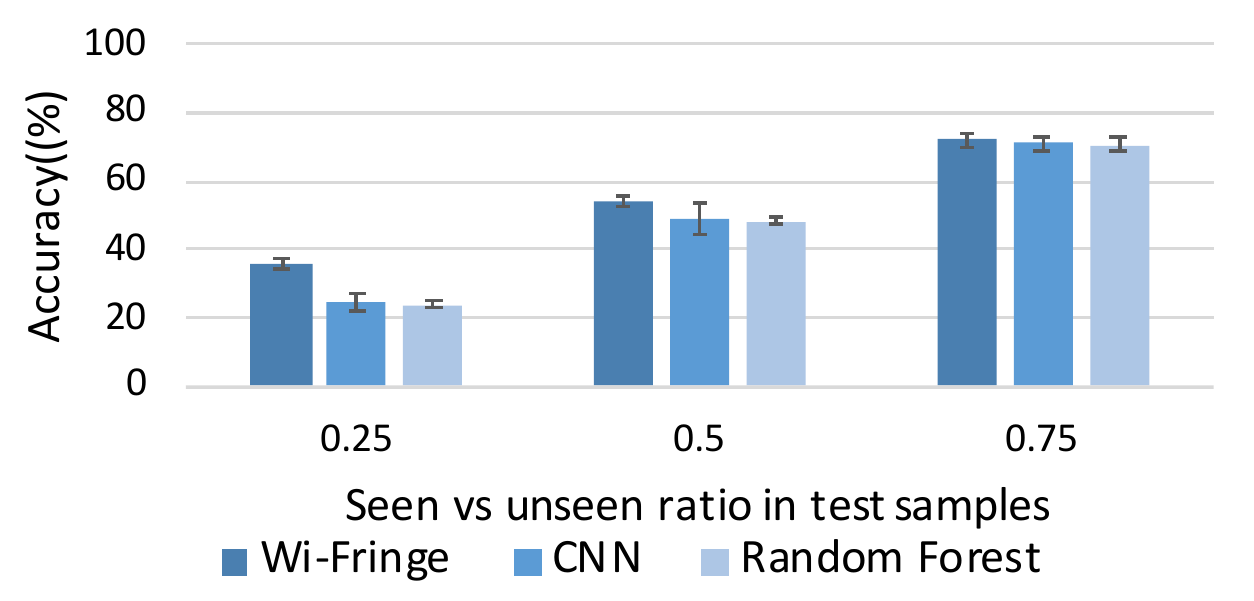} }}%
    \caption{ \Sys performs better than baseline algorithms for all cases. a)When 8 out of 10 classes are in seen category, \Sys has almost $80\%$ accuracy for all cases. On the other hand, baseline algorithms' accuracy drops below $20\%$ for cases when unseen category dominate in test samples. b) For 5 out of 10 classes in seen category \Sys outperforms all baseline algorithms for different cases. c)  When 2 out of 10 classes are in seen class, for $.25$ fraction of test samples coming from seen category \Sys's performance drops below $40\%$ which is still 1.5 times better than baselines' performance.  }
    \label{fig:whole}%
    \vspace{-.5em}
\end{figure*}

To quantify \Sys's end-to-end performance, we report its classification accuracy for an application scenario. We monitor a user's home activity for ten different classes: \mytt{\{push, pull, run, sit, rub, walk, stand, eat, scratch, drink\}}. We consider three different training scenarios. First, we consider that the user provides 8 out of 10 activity classes' examples to \Sys during training, i.e., the number of classes in seen and unseen categories are 8 and 2, respectively. Second, we consider the case where 5 out of 10 activity  classes' examples are given to \Sys during training. The last and the hardest test case is a scenario where \Sys has only 2 activity classes' samples during training, i.e., 8 out 10 classes are unseen. 

In Figure~\ref{fig:whole}, we report the performance of \Sys along with two baseline algorithms: a convolutional neural network (CNN) and a random forest classifier for all three aforementioned scenarios. For each scenario, we consider three cases where we vary the ratio  between  samples from seen and unseen classes in the test dataset in the following ways: a) $\frac{\#seen}{\#unseen}=25\%$, b) $\frac{\#seen}{\#unseen}=50\%$ and c) $\frac{\#seen}{\#unseen}=75\%$. Here, $\#$ denotes number of samples.

In Scenario 1 (Figure~\ref{fig:whole}(a)), where only 2 classes are in the unseen category, \Sys shows an accuracy around $80\%$ for all the cases, whereas the baseline algorithms' accuracy drops below $20\%$ when most of the samples are coming from the unseen category. Note that the unseen classes are chosen by keeping one of their closest neighbours in the word embedding and attribute space in the seen category. 

In scenario 2 (Figure~\ref{fig:whole}(b)), \Sys achieves an accuracy of $84\%$ for case 3 with majority of the samples in test cases coming from the seen classes. However, when the ratio of seen classes in the test data gets decreased in case 1, the accuracy drops to $73\%$. Yet, \Sys's performance is better than both baselines by a margin of greater than $40\%$. 

In scenario 3 (Figure~\ref{fig:whole}(c)),  where only two classes are in the seen category, the accuracy for the case where $75\%$ of test samples are from seen classes reaches up to $72\%$ for \Sys. However, for case 1, the accuracy drops to $36\%$ where $75\%$ of the test samples are from the unseen categories. This drop is due to the fact that most of the classes are now in unseen category and \Sys has very few classes to learn the mapping function from RF to text domain. For this case, baselines achieve a maximum accuracy of only $24\%$. Therefore, it is evident that \Sys's performance is better than traditional classification algorithms in all the cases.

\subsection{Execution Time} Although only a few WiFi chipsets support CSI information extraction, we believe that in the future, with new tools and chipsets, \Sys will be runnable completely on a variety of platforms including smartphones. In this experiment, we report the execution time of the proposed algorithms for a computer (Macbook Pro). Data are read from the file system of the device. In Table~\ref{tab:exec}, we list the execution times for inference of the two models. We see that the state-aware representation (SAR), which has  convolutional and recurrent layers, takes 12ms on average for inference on a Macbook. On the other hand, the execution time for cross-modal projection, which has two fully connected layers, is around 2ms on a Macbook.

\begin{table}[!htb]
\centering
\resizebox{.25\textwidth}{!}{
    \begin{tabular}{r|l}
        \textbf{Network} & \textbf{Macbook Pro} \\
        \hline
        State-Aware Representation & 12ms  \\
        Cross-Modal Projection  & 2ms \\
    \end{tabular}
}

 \caption{Inference time of deep networks.}
 \label{tab:exec}
 	\vspace{-1.5em}
\end{table}

%% file: tex/10_discussion.tex
\section{Discussion}
\label{sec:disc}

\begin{figure}[!htb]
	\centering
	\includegraphics[width = 0.45\textwidth]{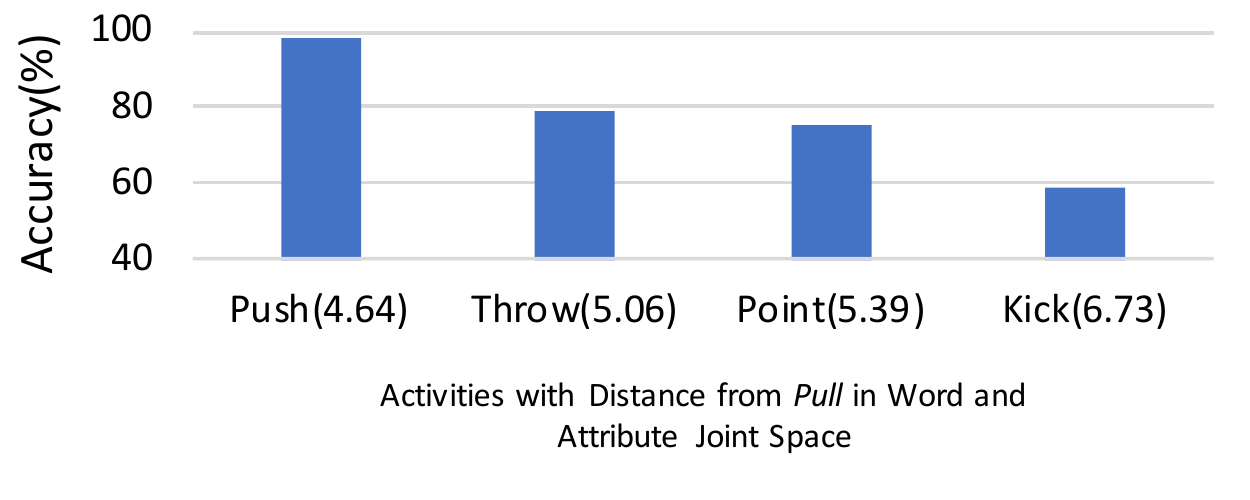}

	\caption{\Sys's performance in unseen activity recognition is dependent on its relationship with classes in seen category.}
	\label{fig:disc}
	\vspace{-1.0em}
\end{figure}

$\bullet$ \textit{Synergey Between Seen and Unseen Class}
 For an unseen activity, \Sys is able to detect it with high accuracy if there is a seen class with similar class label property, i.e., similar word embedding and attribute vectors. \Sys assumes that activity classes with similar properties have semantically similar class labels. We use this assumption to learn the non-linear mapping function to project WiFi signals onto the word embedding and attribute spaces. Without semantically similar training examples corresponding to an unseen class, a zero-shot learner fails to recognize examples from that unseen class. However, in a large dataset, the chances that there is no semantically similar training example to an unseen class is relatively low.  To demonstrate this, we conduct an experiment. We keep \mytt{pull} as an unseen class and exclusively put the following classes in the seen category in a round-robin fashion: \mytt{\{push, throw, point, kick\}}. In Figure~\ref{fig:disc}, on the X-axis, we put these classes with their corresponding distances in the joint word embedding and attribute space with \mytt{pull}. We see that, as the distance increases from the unseen category, the accuracy of unseen class recognition drops. When \mytt{push} is in seen category, the classification accuracy for \mytt{pull} is more than $90\%$. As \mytt{push} and \mytt{pull} are close in word embedding and attribute spaces, the presence of \mytt{push} in training data helps recognize the samples from \mytt{pull}. As we go further away from \mytt{pull} in the word embedding and attribute space, the accuracy drops even more. For \mytt{kick}, which is the farthest from \mytt{pull}, we see that the classification accuracy for \emph{pull} is around $60\%$.

$\bullet$ \textit{Necessity of User-Provided Tag-List.} The user-provided additional labels do not have any influence on the training phase. They are only used in the classification step after the training has been completed. If we do not have these additional labels, then the search space becomes too large (i.e., as big as having all the words in our database for a language) in the classification stage. Therefore, the user provided labels for unseen class is important in getting better accuracy for unseen activities.

$\bullet$ \textit{Environmental Effect.} Our main objective in this paper is to propose the first cross-modal RF to text projection to enable zero-shot activity classification. The proposed representation learning algorithm does not consider environment  or multi-person effect. Recent works~\cite{jiang2018towards,venkatnarayan2018multi} have showed promising results on solving these issues. Both State-Aware Representation and Cross modal projections are modular and generic. We can port these solutions to \Sys to handle these artifacts.

%% file: tex/09_literature.tex
\section{Related Work}
\label{sec:literature}

\subsection{Device-free Sensing} 

$\bullet$ \textit{RSSI based}: WiFi based sensing have opened the doorway for device-free activity monitoring in the last couple of years. Researchers have used wifi signal characteristics such as signal strength (RSSI) and channel state information(CSI) for activity recognition~\cite{kwapisz2011activity}. RSSI based activity recognitions have been proposed in~\cite{abdelnasser2015wigest, sigg2014rf, sigg2014telepathic, kosba2012rasid, sigg2013rf}. However, RSSI based gesture recognition system have limitations in detecting fine-grained gestures. Moreover, all of these works require training examples to detect a particular activity class.~\cite{abdelnasser2015wigest,sigg2014telepathic} uses pattern matching algorithm to find the best match between the target gesture with pre-defined gestures. Our target in this paper, is classifying activities and gestures without training examples and our framework can be ported to RSSI based systems. 

$\bullet$ \textit{Special Device based}: There have been several works in gesture and activity recognition using specialized devices and radars.~\cite{malysa2016hidden, yue2018extracting, zhao2016emotion} use FMCW~\cite{ramasubramanian2017using, adib2015smart} radio to monitor user activity.~\cite{yue2018extracting} trains a CNN classfier to detect human motion.~\cite{malysa2016hidden} trains a hidden markov model using time-velocity feature for activity recognition. \cite{pu2013whole} uses a 5-antenna receiver and a single-antenna transmitter to perform gesture classification, in the presence of three other users performing random gestures. 
They use doppler shifts to match with pre-defined gestures for an incoming test data. It is evident that none of these specialized device based sensing system deals with ctivity recognition without prior examples. They all require labelled training examples for activity recognition. 

$\bullet$ \textit{CSI based}: With the availability of CSI from network interface cards, multiple works~\cite{wang2017wifall, wang2016we, xi2014electronic, li2016wifinger} have emerged which exploit CSI information for gesture and activity recognition. \cite{wang2014eyes} proposes a signal profile matching technique to detect loosely defined daily activities that involve a series of body movements over a certain period of time.~\cite{wang2015understanding} proposes correlation  between CSI amplitude value and gesture speed to build model for gesture recognition.  ~\cite{wang2016gait} uses variations in the Channel State Information (CSI) to classify gaits of humans. ~\cite{virmani2017position} uses translation based data augmentation technique to make gesture classification models robust to user orientation.~\cite{venkatnarayan2018multi} proposed multi-person gesture recognition system by generating virtual gesture samples and combining them to create an exhaustive template matching algorithm. Recent works such as ~\cite{chen2018wifi, wang2018spatial, ma2016survey} use deep learning based techniques such  Convolutional Neural Networks to recognize activities from CSI.~\cite{zou2018deepsense, chen2018wifi} proposes recurrent neural network based activity classifier , however ours is the first work to model the micro-activity or state transition which constitute an activity.~\cite{zou2018deepsense}  predicts label for each segment of the CSI stream. On the other hand, we need only one label for the whole CSI stream which makes our model to learn the entire sequence of states to make a decision about the activity. Besides, we learn the local and transitional feature in an end to end manner with bi-directional recurrent neural network, which makes our model stronger. Ours  SAR is the first model to incorporate local feature of states and their transition in a bi-directional fashion. ~\cite{jiang2018towards} proposed an adversarial network to learn environment independent signal characteristics from gestures. In this work, the authors used unlabelled data to improve their model's performance. But their proposed system is not able to infer an activity without having any training example. Very recent work~\cite{zero_effort} proposed velocity profiles as environment independent feature to solve the problem of environmental effect on CSI. All of these works rely on provided training examples to classify a particular class of activity. \Sys deals with classification of activity from WiFi CSI data without any training examples. This is significantly different from current state of the arts.

\subsection{Zero Shot Learning}
In image domain, learning from limited or no data has been explored recently.~\cite{koch2015siamese, vinyals2016matching, duan2017one} focus on learning suitable image representation to classify images from very few or only one examples. These works are totally different from ours as we want to recognize activities from WiFi without any training data.
The other branch of learning with limited data focuses on zero shot learning for image classification.
The  earlier practices of zero shot learning~\cite{lampert2014attribute,norouzi2013zero, kodirov2015unsupervised} for image classification problem infer the labels of unseen classes using a two step algorithm. First, the attributes of the sample is inferred and then the class label is predicted from an attribute database, which has most similar attributes with the image. Recent works~\cite{xian2016latent, socher2013zero, akata2016label} have explored the mapping between image features and semantic space by projecting image features into word embedding space. Similar line of works~\cite{palatucci2009zero,frome2013devise, zhang2015zero,kodirov2017semantic} project image feature into semantic space then searches the nearest class label embedding using ranking loss. Although these papers propose zero shot learning method for images, none of them addresses the problem for RF domain and activity recognition. ~\cite{zellers2017zero, xu2017transductive,liu2017generalized, madapana2017semantical} do activity recognition using zero shot learning for RGBD data.~\cite{islam2019soundsemantics} proposes zero shot learning for audio event recognition. But, the sequential nature of activity and need for external attribute knowledge makes our problem more challenging. ~\cite{cheng2013towards} proposes zero shot learning for body-worn IMU based activity recognition. However, our work is the first paper to propose a zero shot learning method for WiFi based activity classification where we overcome the challenges for cross-modal learning between text and RF domain.  These works on zero shot activity detection use an attribute database of the gestures to recognize unseen types. Our contextual word-embedding based approach does not require the attributes of the unknown gestures or activities beforehand. To the best of our knowledge, we present the first system with capabilities of inferring activities from RF signal without training examples. We propose the first work to integrate textual domain with RF. We also present novel feature representation for RF based activity monitoring.

%% file: tex/11_conclusion.tex
\section{Conclusion}
\label{sec:conclusion} 
In this paper, we present the first WiFi-based device-free activity recognition system that does not require training examples for all activities. We propose a robust state-aware representation of RF signature associated with activities to preserve their contextual information. We propose a novel way to embed contextual information from the text domain to the RF domain by projecting RF data onto the word embedding and attribute space. We use this cross-modal RF embedding and propose a general classifier to recognize both \emph{seen} and \emph{unseen} activities. We collect WiFi data for 20 different activities from four volunteers and show that \Sys is capable of inferring activities from WiFi without training examples with $62\%-90\%$ accuracy for 2--6 unseen classes. 
\newpage